\documentclass[12pt,epsf]{article}
\usepackage{verbatim}

\usepackage[english]{babel}

\usepackage{amssymb,amsmath}
\usepackage{graphicx, xcolor, varwidth}
\usepackage{setspace}
\usepackage[permil]{overpic}
\usepackage{cite}

\usepackage{dsfont}

\colorlet{darkblue}{blue!70!black}
\newcommand{\arXiv}[1]{\href{http://www.arXiv.org/abs/#1}{#1}}
\usepackage[colorlinks=true,urlcolor=darkblue,linktocpage=true,linkcolor=darkblue,citecolor=darkblue]{hyperref}

\numberwithin{equation}{section}

\newcommand{\be}{\begin{equation}}
\newcommand{\ee}{\end{equation}}
\newcommand{\bea}{\begin{eqnarray}}
\newcommand{\eea}{\end{eqnarray}}
\newcommand{\bear}{\begin{eqnarray}}
\newcommand{\eear}{\end{eqnarray}}
\newcommand{\beas}{\begin{eqnarray*}}

\newcommand{\eeas}{\end{eqnarray*}}
\newcommand{\ba}{\begin{array}}
\newcommand{\ea}{\end{array}}

\newcommand{\tr}{\operatorname{tr}}
\newcommand{\pd}[2][1]{\ifnum#1=1 \frac{\partial}{\partial {#2}} \else
  \frac{\partial^#1}{\partial {#2}^{#1}}\fi}
\newcommand{\dpd}[2][1]{\ifnum#1=1 \dfrac{\partial}{\partial {#2}} \else
  \frac{\partial^#1}{\partial {#2}^{#1}}\fi}
\newcommand{\td}[2][1]{\ifnum#1=1 \frac{d}{d{#2}} \else
  \frac{d^#1}{d{#2}^{#1}}\fi}

\renewcommand{\(}{\left(}
\renewcommand{\)}{\right)}

\newcommand{\nbox}{{\,\lower0.9pt\vbox{\hrule \hbox{\vrule height 0.2 cm \hskip 0.19 cm \vrule height 0.2 cm}\hrule}\,}}
\newcommand{\Tr}{\ {\rm Tr}\ }

\newcommand{\eps}{\epsilon}

\newcommand{\ie}{{\it i.e.,}\ }

\newcommand{\bL}{\mathbf{L}}

\newcommand{\btau}{\bar{\tau}}
\renewcommand{\bL}{\overline{L}}

\newcommand{\bq}{\bar{q}}
\newcommand{\bh}{\bar{h}}
\newcommand{\bx}{\bar{x}}
\renewcommand{\c}[1]{\frac{c}{#1}}
\newcommand{\half}{\tfrac{1}{2}}

\newcommand{\bz}{\bar{z}}
\newcommand{\tsigma}{\tilde{\sigma}}

\newcommand{\bw}{\bar{w}}

\newcommand{\by}{\bar{y}}



\newcommand{\AdS}{\text{AdS}}
\newcommand{\CFT}{\text{CFT}}
\newcommand{\beq}{\begin{equation}}
\newcommand{\eeq}{\end{equation}}

\begin{document}
\begin{spacing}{1.3}
\begin{titlepage}

\begin{center}
{\Large \bf Entanglement Scrambling \\
\vspace{.5cm}
in 2d Conformal Field Theory}

\vspace*{15mm}

Curtis T.~Asplund$^*$, Alice Bernamonti$^\ddag$,  Federico Galli$^\ddag$, and Thomas Hartman$^\S$
\vspace*{6mm}

\textit{$^*$ Department of Physics, Columbia University, 538 West 120th Street, New York, New York 10027, USA}\\ 
\textit{$^\ddag$ Instituut voor Theoretische Fysica, KU Leuven, Celestijnenlaan 200D,\\ B-3001 Leuven, Belgium}\\
\textit{$^\S$ Department of Physics, Cornell University, Ithaca, New York 14853, USA}\\ 

\vspace{6mm}

{\tt ca2621@columbia.edu, alice@itf.fys.kuleuven.be, federico@itf.fys.kuleuven.be, hartman@cornell.edu}

\vspace*{6mm}
\end{center}
\begin{abstract}

We investigate how entanglement spreads in time-dependent states of a 1+1 dimensional conformal field theory (CFT).  The results depend qualitatively on the value of the central charge.  In rational CFTs, which have central charge below a critical value, entanglement entropy behaves as if correlations were carried by free quasiparticles. This leads to long-term memory effects, such as spikes in the mutual information of widely separated regions at late times. When the central charge is above the critical value, the quasiparticle picture fails.  Assuming no extended symmetry algebra, any theory with $c>1$ has diminished memory effects compared to the rational models.  In holographic CFTs, with $c \gg 1$, these memory effects are eliminated altogether at strong coupling, but reappear after the scrambling time $t \gtrsim \beta \log c$ at weak coupling.

\end{abstract}

\end{titlepage}
\end{spacing}

\vskip 1cm

\tableofcontents

\begin{spacing}{1.3}

\section{Introduction}

How does quantum information spread in a strongly coupled system, far from equilibrium?  This question is central to the dynamics of a wide variety of experimental and theoretical systems, from cold atoms and condensed matter to quantum chaos and black holes (see for example \cite{bloch},\cite{cm},\cite{Sekino:2008he} and \cite{holoq,Hartman:2013qma}). 

Quantum information is encoded in entanglement, so a useful way to characterize its spread is through the time-dependent entanglement entropy of spatial subsystems.  In this paper, we investigate this quantity in 1+1d critical systems, in time-dependent states which have only short-range correlations at time $t=0$.  These states model a global quench from a gapped Hamiltonian.

The entanglement entropy of a single, connected region in this context was computed by Calabrese and Cardy using conformal field theory (CFT) \cite{Calabrese:2005in,Calabrese:2007rg,Calabrese:2009qy}.  In the limit where the time $t$ and size $L$ of the interval are much larger than the initial correlation length $\xi$, the result is fixed universally by conformal symmetry, and depends only on the central charge $c$ of the CFT. It is equal to $c$ times the answer for a free boson. This can be modeled by assuming that entanglement is carried by free quasiparticles, propagating ballistically at the speed of light.

The entanglement entropy of disjoint regions is a more detailed probe of how information spreads.
We focus on the case of two disjoint intervals $A$ and $B$. 
In this case the mutual information $I(A,B) = S_A +S_B - S_{A\cup B}$ bounds the strength of connected correlation functions of smeared local operators in the two regions \cite{MIbound}. Our main result is that the entanglement entropy $S_{A\cup B}(t)$ is not universal in general, but depends qualitatively on the value of the central charge: It interpolates between the free quasiparticle answer for small $c$, and the holographic prediction of three-dimensional quantum gravity at large $c$.


\subsubsection*{Free quasiparticles vs.~quantum gravity}

\begin{figure}
\begin{center}
\includegraphics[width=0.95\textwidth]{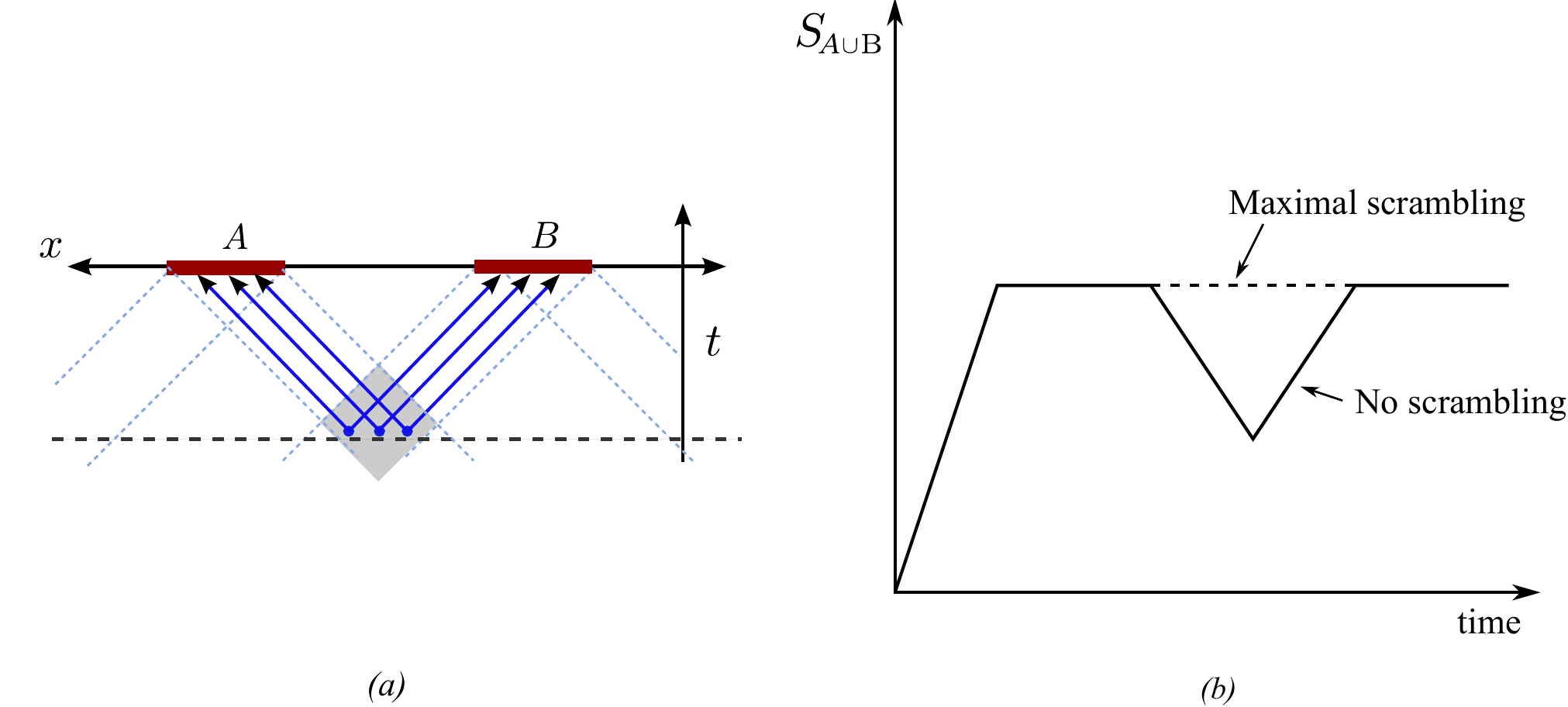}
\caption{\small $(a)$ Quasiparticle picture for the entanglement entropy of two disjoint intervals. Entangled pairs are created at $t=0$, which propagate in opposite directions at the speed of light. When one of the particles enters $A$ and the other enters $B$, the entanglement entropy $S_{A\cup B}$ decreases. $(b)$ Memory effect in the entanglement entropy of two separated intervals after a global quantum quench.  For intervals of size $L$ separated by a distance $D>L$, the dip is centered at $t = (D+L)/2$ and extends over the range $D/2 < t < D/2 + L$.  The solid line is the free quasiparticle answer, and the dashed line is the holographic answer. \label{fig:dip}}
\end{center}
\end{figure}

Suppose the regions $A$ and $B$ both have length $L$ and are separated by a distance $D>L$, as in figure \ref{fig:dip}$a$. In a theory of free quasiparticles, the entanglement entropy $S_{A\cup B}(t)$ is shown in figure \ref{fig:dip}$b$.  After the quench, it grows linearly as entangled pairs spread, and saturates at the thermal value for temperature $\beta \sim \xi$.  At $t = D/2$, entangled pairs created in the middle enter the two regions, as shown in figure \ref{fig:dip}$a$, leading to a dip in $S_{A\cup B}$ and a corresponding spike in the mutual information.

At the midpoint of the dip, the left-moving quasiparticles in region $A$ are maximally entangled with the right-moving quasiparticles in region $B$, so $S_{A\cup B}$ is exactly half the thermal value (after subtracting the divergent contribution at $t=0$). In this scenario, maximally entangled degrees of freedom remain maximally entangled even as they propagate to very large separation.  In other words, entanglement does not scramble.

It was claimed in \cite{Calabrese:2005in,Calabrese:2009qy} that the free quasiparticle behavior for multi-interval entanglement after a global quench is universal to all conformal field theories. This is surprising in a strongly coupled system, and we will argue that an assumption underlying this claim (about the analytic continuation of certain Euclidean correlators to Lorentzian signature) is justified only in a restricted class of theories with a large degree of symmetry. A first hint of this was provided by holographic calculations of entanglement entropy  \cite{Asplund:2013zba,Balasubramanian:2011at,Allais:2011ys}, where, even in 1+1 dimensions, entanglement scrambles \textit{maximally} --- late time features in the entanglement entropy of widely separated subsystems, such as the dip in figure \ref{fig:dip}$b$, are entirely absent. This apparent discrepancy was first observed in   \cite{Asplund:2013zba} and has also been explored in \cite{Leichenauer:2015xra}. 
Our aim is to reconcile these two pictures, and to understand the middle ground.  

\subsubsection*{Results}
We argue that, in fact, the scrambling of entanglement is generic in 1+1d conformal field theory: scrambling occurs whenever the central charge is greater than a critical value depending on the number of conserved currents.  In a CFT without extended chiral symmetry, the stress tensor and its descendants are the only conserved currents, and entanglement scrambles if and only if $c>1$. More generally, we define an effective central charge $c_{\rm currents}$ for the chiral sector of the theory, and find that entanglement scrambles when $c > c_{\rm currents}$.  

The quasiparticle picture is accurate in theories with $c = c_{\rm currents}$.  These include the rational CFTs, as well as some non-rational theories like the compact boson at irrational radius squared.  We refer to this class of theories as current dominated, for the following reason. Denoting the density of states by $\Omega(h,\bh)$, where $(h,\bh)$ are the left and right conformal weights, $c = c_{\rm currents}$ implies
\be\label{fact}
\Omega(h,\bh) \approx \Omega(h,0)\times \Omega(0, \bh) \quad \mbox{as} \quad h,\bh \to \infty \ .
\ee
That is, the spectrum of high-dimension operators is dominated by conserved currents, which have dimension $(h,0)$ or $(0,\bh)$. Correlators in such theories also factorize into left and right moving contributions, in a stronger sense than in general CFTs. 

The opposite extreme is a holographic CFT, which has $c_{\rm currents} \sim 1$ 
and $c \gg 1$, and scrambles maximally. We recover this universal answer directly from our CFT analysis in the holographic limit. Theories with $c>c_{\rm currents}$, but where $c$ is not large, also scramble, but the entanglement entropy in these theories is not universal.  These intermediate theories are not rational, so they cannot be solved exactly, and perhaps for this reason have received much less attention in the literature than rational CFTs.  They are, however, perfectly normal, unitary quantum field theories, with a moderate central charge, a unique normalizable vacuum state, and a discrete spectrum when placed in finite volume. Such CFTs with $c>c_{\rm currents}$ are much like their higher-dimensional cousins, since rational CFTs do not exist in $d>2$. They have an infinite but discrete set of primary fields. An example is a sigma model with a compact target space that has no isometries (or few isometries). Such a theory does not have the large quasiparticle dip shown in fig.~\ref{fig:dip}, even in the scaling limit where times and lengths are much greater than the initial correlation length. An explicit example with $c=12$ is discussed in section~\ref{s:discussion}.

The evidence for these claims is strong but not entirely conclusive.  It follows from calculations of the R\'enyi entropies $S_n$  in conformal field theory, which are related to the entanglement entropy by extrapolating $n \to 1$.  We consider two models for a quantum quench: the boundary state model of Calabrese and Cardy \cite{Calabrese:2005in,Calabrese:2007rg,Calabrese:2009qy}, and the thermal double model of Hartman and Maldacena \cite{Hartman:2013qma}. In the thermal double model, we show that the $S_2$ R\'enyi entropy behaves precisely as claimed, and argue (somewhat incompletely) that the same applies for all $S_n$.  Thus the evidence for scrambling in this model is very strong.  In the boundary state model, our conclusions hold as long as boundary states in non-rational conformal field theory do not, for some unknown reason, produce a spurious singularity in correlation functions, beyond the usual singularities required by the operator product expansion, and the additional singularities from image points across the boundary.  This assumption is unproven but seems likely without evidence to the contrary, so we suspect that our conclusions hold also for boundary states. We therefore disagree with the claim in \cite{Calabrese:2005in,Calabrese:2009qy,Coser:2014gsa} that the quasiparticle picture for multiple-interval entanglement entropy after a quench is entirely universal --- the derivation implicitly assumes that the theory is current dominated. The differences appear only for multiple intervals after a quench, or certain configurations of highly boosted intervals in vacuum, so the results for a single interval are unaffected. Similar differences would appear in the calculation of $n$-point correlation functions of local operators after a quench \cite{Calabrese:2006rx}, but only for $n>2$. 

We emphasize that our conclusions hold in the scaling limit where length and time scales are much larger than the correlation length in the initial state (as in \cite{Calabrese:2005in,Calabrese:2007rg,Calabrese:2009qy}). The scaling limit is taken with $c$ held fixed, and we assume $c$ is large only when stated explicitly in section \ref{s:largec}.  In holographic theories, there is an additional consideration since we must take both the large-$c$ limit and the scaling limit.  This order of limits is not the origin of the discrepancy between holographic and rational CFTs. The interplay between the scaling limit and the large-$c$ limit is addressed in section \ref{ss:stringy}. At strong coupling, assuming a sparse spectrum of low-dimension operators, the conformal field theory answer reproduces the holographic prediction for mutual information after a global quench. In weakly coupled CFT, it leads to a quantitative prediction for stringy corrections to the holographic entanglement result.

\subsubsection*{Outline}
The Calabrese-Cardy boundary state model for a global quench, and the related technology of twist correlation functions in CFT, is reviewed in sections~\ref{sec:bdrystate}-\ref{sec:eebdry}.  In section~\ref{ss:lcdip}, we show that around the time of the dip in the quasiparticle entanglement entropy, the behavior of the R\'enyi entropies is controlled by a light cone singularity in the twist correlator, and that this singularity is not fixed by the ordinary operator product expansion.

The aim of section~\ref{sec:renyi} is to understand the nature of this light cone singularity.  First, we relate it to a technically simpler question about the mutual information of offset intervals in a thermal double CFT.  In this setting, in section~\ref{sec:2ndRenyi} we compute the singularity in the $S_2$ R\'enyi entropy by conformally mapping the replica geometry to the torus partition function and give a definition of $c_{\rm currents}$. 
Based on properties of the spectrum in rational (or nearly rational) vs non-rational CFT, we conclude that the quasiparticle picture produces the correct $S_2$ only for central charge $c$ below a critical value.  In section~\ref{ss:cbe} we attempt to extend this to general R\'enyi index $n$, but only partially succeed. Section~\ref{ss:c=1} is devoted to the marginal case of the $c=1$ free boson, and the question of whether the dip gets larger or smaller is addressed in section~\ref{ss:dipbound}.

In section \ref{s:largec} we explore the large $c$ holographic limit.  The gravity calculations of mutual information and second R\'enyi entropy after a global quench are sketched in sections~\ref{sec:HM}-\ref{ss:gravityren}; these are reproduced from CFT (with some extra assumptions) in~\ref{sec:CFTlargec}; and the weakly coupled limit of a particular holographic CFT, the D1-D5 symmetric orbifold, is considered in section~\ref{ss:stringy}.

In the conclusions, we briefly comment on several open questions and possible extensions of these results, including local quenches and entanglement negativity.

\section{Origin of entanglement memory}\label{s:origin}

Consider a 1+1d CFT, in an initial pure state $|\Psi\rangle$ with a finite correlation length $\xi$ and finite energy density.    This state may be produced, for example, by globally quenching the Hamiltonian of a gapped system to a critical point.  At time $t=0$, regions of size $L \gg \xi$ exhibit area-law entanglement.  The state is time dependent, and since it has finite energy density, we may expect it to exhibit volume-law entanglement for a single interval at late times.  This was beautifully demonstrated in \cite{Calabrese:2005in,Calabrese:2007rg,Calabrese:2009qy}.  

A more detailed probe of entanglement in this system is the entanglement entropy or mutual information of separated intervals.  Define the regions
\be\label{defab}
A:\, x \in [0, L] , \qquad B: \, x \in [D+L, D+2L] \ .
\ee
To simplify the discussion we will always assume 
\be
t > 0 , \qquad D>L \,,
\ee
and only comment on the case $D<L$ in the conclusions.

The entanglement entropy $S_{A\cup B}$ does not measure the entanglement of $A$ with $B$, but rather the entanglement of $A \cup B$ with the rest of the system --- if $A$ and $B$ are highly entangled with each other, then they cannot be highly entangled with degrees of freedom elsewhere, so $S_{A\cup B}$ is small. It is UV divergent, but the divergence is time-independent, and could be removed entirely by considering instead the mutual information
\be
I(A,B) = S_A + S_B - S_{A \cup B} \ .
\ee
Two possible behaviors of the entanglement entropy as a function of time are illustrated in figure \ref{fig:dip}. The linear rise at early times comes from correlations spreading until $t \sim L/2$.  The dip, when it exists, can be understood as coming from entangled degrees of freedom that started at the midpoint between the two intervals, with left-movers entering region $A$ and right-movers entering region $B$ around $t \sim D/2$. When there is a dip in $S_{A\cup B}$, there is a corresponding peak in the mutual information.

In this section we will setup the calculation of the entanglement entropy, following \cite{Calabrese:2005in,Calabrese:2007rg,Calabrese:2009qy}, and isolate the part of the computation responsible for the dip around $t \sim D/2$.  

\subsection{Review of the boundary state model}\label{sec:bdrystate}
The experiment just described is in a particular pure state, $|\Psi\rangle$.  We cannot hope to capture the exact details of an arbitrary pure state, but it was argued by Calabrese and Cardy that the behavior of the entanglement entropy (and correlation functions) for $t, L, D \gg \xi$ is universal.  The universal features of $|\Psi\rangle$ are captured by the state
\be\label{bstate}
|\Phi\rangle = e^{-\beta H/4} |{\cal B}\rangle
\ee
where $|\cal{B}\rangle$ is a conformal boundary state, $H$ is the CFT Hamiltonian, and $\beta \sim  \xi$. The boundary state $|\cal{B}\rangle$ is not normalizable, but evolution in Euclidean time spreads out correlations and renders the state normalizable.  

Entanglement entropy in the state $|\Phi\rangle$ can be computed by the replica method.  Define the R\'enyi entropy
\be
S_n = \frac{1}{1-n}\log \Tr \rho^n \ ,
\ee
where $\rho$ is the reduced density matrix of a subregion, and $n \geq 2$ is an integer.  
The entanglement entropy $S = -\tr \rho \log \rho $ is obtained by analytic continuation in $n \to 1$.

\subsubsection*{Twist correlators}
In the state \eqref{bstate}, the R\'enyi entropy can be computed by a path integral.  For the region $A\cup B$ defined above, matrix elements of $\rho_{A\cup B}$ at $t=0$ are computed by the path integral on an infinite strip:
\be
\langle \varphi' | \rho_{A \cup B} | \varphi\rangle = \begin{gathered} \includegraphics{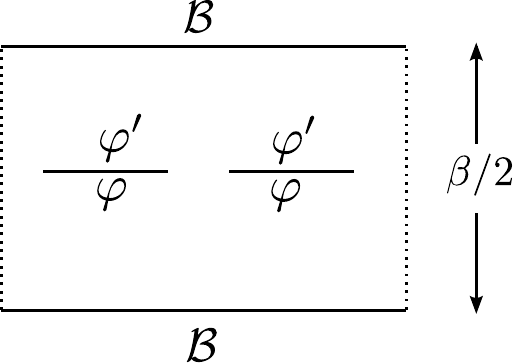} \end{gathered} \ ,
\ee 
where $\varphi$ and $\varphi'$ denote states on region $A \cup B$. Reading this diagram from bottom to top, the boundary condition at the bottom defines the state $|\cal{B}\rangle$, the Euclidean time evolution over the bottom half of the strip produces $e^{-\beta H/4}|\cal{B}\rangle$, and the path integral over the upper half corresponds to $\langle {\cal B} | e^{-\beta H/4}$. 

The replica partition function $\Tr \rho_{A\cup B}^n$ is computed by the path integral on $n$ copies of this system, glued along region $A \cup B$:
\be\label{gluing}
\Tr \rho_{A\cup B}^n = \begin{gathered}\includegraphics{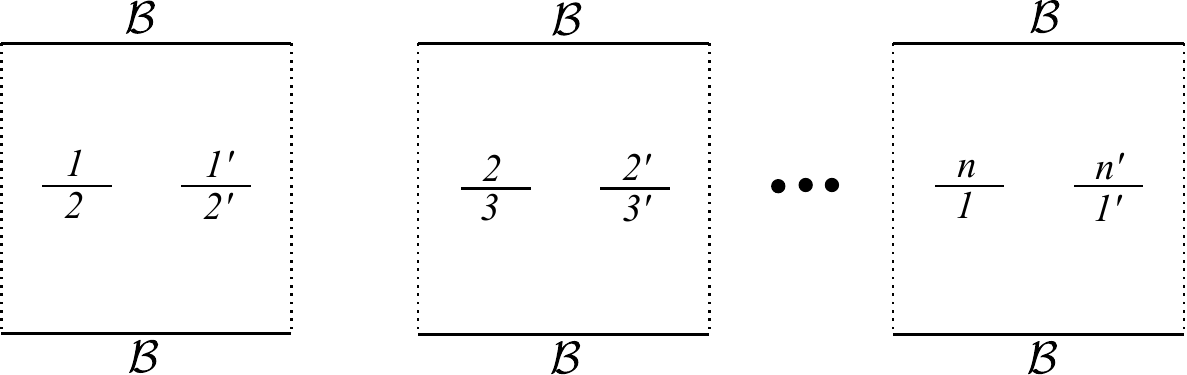}\end{gathered} \ ,
\ee
with slits identified as indicated by the numbers $1\dots n$, $1'\dots n'$. This is a path integral on a Riemann surface with $n-1$ handles, and $n$ holes where we impose boundary condition ${\cal B}$. It can also be viewed as a correlation function of twist operators on a single sheet,
\be\label{twistcor}
\Tr \rho_{A\cup B}^n  \sim I_n^{\text{strip}} \equiv  \langle \sigma(z_1, \bz_1) \tsigma(z_2, \bz_2)  \sigma(z_3, \bz_3) \tsigma(z_4, \bz_4) \rangle_{\text{strip}} \ ,
\ee
where $(z_1, \bz_1)$ and $(z_2, \bz_2)$ are the endpoints of region $A$, and $(z_3, \bz_3)$ and $(z_4, \bz_4)$ are the endpoints of region $B$.   This correlator is computed on a strip of height $\beta/2$, with boundary condition $\cal{B}$ at the top and bottom. The twist operators $\sigma$ and $\tsigma$, which have opposite orientation, are defined to reproduce the multisheeted path integral in \eqref{gluing}. This definition, together with the conformal transformation properties of the path integral, lead to the identification of the conformal weights of the twist operators $(L_0, \bL_0) = (h_n, h_n)$ as  \cite{Holzhey:1994we,Calabrese:2004eu}
\be
h_n = \frac{c}{24}(n-1/n) \ .
\ee
The twist operators are not local operators in the original CFT, but they are local operators in the orbifold $\CFT^n/\mathbb{Z}_n$, so we can use standard methods to compute these correlators in the orbifold theory.

For Im $z_i = \frac{\beta}{4}$, with $\bz_i=z_i^*$, the correlator \eqref{twistcor} computes the replica partition function at $t=0$.  To calculate the R\'enyi entropy as a function of real time, we first compute the correlation function for Euclidean insertions off the real line with arbitrary $z_i$, then analytically continue to Lorentzian signature by taking $z$ and $\bz$ to be independent complex numbers.
For the configuration \eqref{defab}, 
\begin{align}
z_1 &= -t +i\beta/4,&  \quad \bz_1 &= t - i \beta/4\\
z_2 &= L-t+i\beta/4 , & \quad \bz_2 &= L+t-i\beta/4\notag\\
z_3 &= D+L-t +i\beta/4, &\quad \bz_3 &= D+L+t-i\beta/4\notag\\
z_4 &= D+2L-t +i\beta/4, &\quad \bz_4 &= D+2L+t-i\beta/4 \notag
\end{align}

\subsubsection*{Mapping to the upper half plane}
\begin{figure}
\centering
\includegraphics{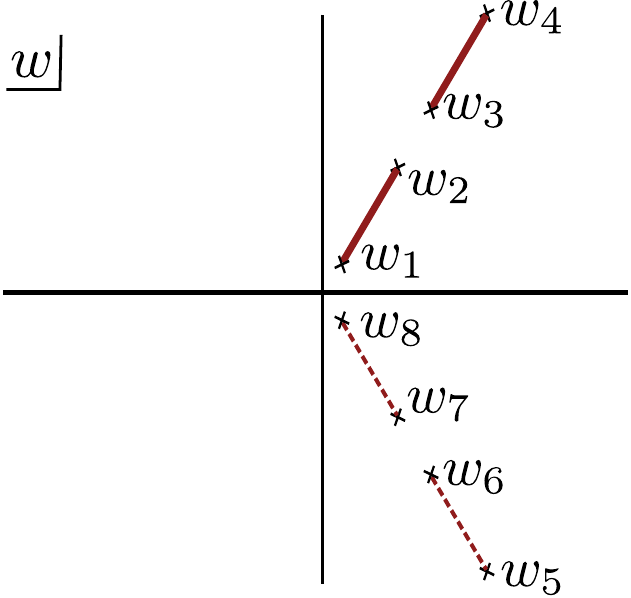}
\caption{Schematic illustration of the method of images. The conformal transformation properties of the correlator with insertions $w_{1,2,3,4}$ in the UHP are identical to those of the holomorphic 8-point function in the full plane, with image points $w_{i=5,6,7,8} = \bw_{9-i}$. The solid red lines indicate the regions $A$ and $B$. \label{fig:images}}
\end{figure}

The correlation function on a strip  \eqref{twistcor}, with boundary condition ${\cal B}$ at the top and bottom, can be conformally mapped to a correlator on the upper half plane (UHP) with boundary condition ${\cal B}$ on the real line, via
\be
w = e^{2\pi z/\beta} \ .
\ee
The correlator is then
\be\label{strip}
I_n^{\text{strip}} = \left(2\pi \over \beta\right)^{8h_n} |w_1 w_2 w_3 w_4|^{2h_n} I_{n}^{\text{UHP}}(w_i, \bw_i) \ ,
\ee
where
\be
I_{n}^{\text{UHP}} \equiv \langle \sigma(w_1, \bw_1) \tsigma(w_2, \bw_2) \sigma(w_3, \bw_3) \tsigma(w_4, \bw_4)\rangle_{\text{UHP}} \ .
\ee
In general, $I_n^{\text{UHP}}$ is not fixed by conformal invariance and cannot be computed.  However its properties under conformal transformations can be understood from the method of images:  The conformal Ward identity for a 4-point function in the UHP is identical to the conformal Ward identity for a holomorphic 8-point function in the full plane, where for each insertion $w_i$ in the UHP we include an image point $w_{i=5,6,7,8} = \bw_{9-i}$ in the lower half-plane \cite{BCFT}. That is, $I_n^{\text{UHP}}(w_{1,2,3,4}, \bw_{1,2,3,4})$ behaves under conformal transformations exactly like an 8-point function
\be
\langle \phi(w_1) \phi(w_2) \phi(w_3) \phi(w_4) \phi(w_5) \phi(w_6) \phi(w_7) \phi(w_8)\rangle_{\text{plane}}
\ee
with the image points
\be
w_5 = \bw_4 \ , \quad  w_6 = \bw_3\ ,  \quad w_7 = \bw_2 , \quad w_8 = \bw_1 \ .
\ee
This is illustrated in figure \ref{fig:images} (for $t$ purely imaginary). We have written $\phi$ rather than $\sigma$ in this correlator because the 4-point function on the UHP is not actually equal to an 8-point function of twist operators on the full plane, it just has the same transformation properties. There is not necessarily an actual (local) operator $\phi$ in the theory whose 8-point function reproduces $I_{\text{UHP}}$.


\subsection{Entanglement entropy in the boundary state model}\label{sec:eebdry}
The time-dependent entanglement entropy in this model has been discussed many times in the literature \cite{Calabrese:2005in,Calabrese:2007rg,Calabrese:2009qy,Coser:2014gsa}. However, one of our goals is to understand exactly when those results apply, so we will repeat the calculation in detail. We begin with a discussion of the regimes away from the dip, where we agree with the literature, then return to the possibility of a dip in the next subsection.

The reason we can make further progress in the scaling limit $t, L, D \gg \beta$ is that, in some cases, the correlator in this limit is fixed by the short-distance behavior of the twist operators. There are two limits where this short distance behavior is known \cite{Calabrese:2004eu,Calabrese:2009qy,Headrick:2010zt}.  First, a twist operator $\sigma$ in the UHP may approach another twist operator $\tsigma$ in the UHP.  In this case we can ignore the boundary, and the short-distance behavior is
\be\label{limita}
\sigma(x, \bx)\tsigma(y, \by) \sim (x-y)^{-2h_n} (\bx-\by)^{-2h_n}  \qquad (\mbox{as\ } x \to y\ , \bx \to \by) \ .
\ee
Second, a twist operator in the upper half plane may approach the boundary.  It is therefore approaching its image point, and the short distance behavior is
\be\label{limitb}
\sigma(x, \bx) \sim r_n (x - \bx)^{-2h_n} \qquad (\mbox{as\ } x \to \bx) \ ,
\ee
where $r_n$ is a constant (since we have already fixed the normalization of the operator in \eqref{limita}).

In Euclidean signature, \eqref{limita} and \eqref{limitb} are the only types of short distance limit. The singularities in Lorentzian signature are more complicated, since $w$ and $\bw$ are independent; we will return to this below. 

The two Euclidean limits fix the behavior of the Lorentzian correlator $I_n^{\text{UHP}}$ in certain regimes where all four twist operators are in one of a limits \eqref{limita} or \eqref{limitb}.  To see when this occurs, we will look at the cross ratios
\be
\eta_{ij} = \bar{\eta}_{ij} = \frac{(w_i - w_j)(\bw_j - \bw_i)}{(w_i-\bw_j)(w_j-\bw_i)} 
\ee
in different regimes. Only five of these six real cross ratios are independent, but it is useful to keep track of all of them.

\subsubsection*{Early times}
At early times,
\be\label{earlyrange}
0 < t < \frac L 2 \ ,
\ee 
the $\beta \to 0$ limit sends
\be
\eta_{ij} \to 1 \ .
\ee
This is a configuration where each twist operator approaches its image point, $w_{i} \to \bw_i$.  This is not immediately obvious, since there are multiple simultaneous limits. For example, 
\be
\frac{w_2 - w_1}{w_2 - \bw_2} \to 0 \ ,
\ee 
but this does not contradict $w_2 \to \bw_2$, as  $\sigma(w_1, \bw_1)$ goes to the boundary faster than to any other operator.
That is, $w_1$ approaches $\bw_1$ without any other operators nearby and this means we can apply the boundary OPE \eqref{limitb}, $\sigma(w_1,\bw_1) \sim r_n(w_1 - \bw_1)^{-2h_n}$. Having eliminated $\sigma(w_1, \bw_1)$, we can then repeat the same argument for $w_2 \to \bw_2$, then for $w_3 \to \bw_3$, and then for $w_4 \to \bw_4$. 
Therefore the leading term in the correlator is
\bea
I_n^{\text{UHP}} &\sim& r_{n}^4 \left[ (w_1 - \bw_1)(w_2 - \bw_2) (w_3 - \bw_3) (w_4 - \bw_4)\right]^{-2h_n} \\
&\sim& r_{n}^4  \exp\left[ -\frac{8\pi}{\beta}h_n(D+2L+2t ) \right] \ .\notag
\eea
Plugging into \eqref{strip},
\be
I_n^{\text{strip}} =  r_{n}^4 \left( 2\pi \over \beta\right)^{8 h_n}\exp\left[ -\frac{16 \pi}{\beta}h_n  t \right] \ .
\ee
The resulting entanglement entropy in the scaling limit is
\be
S_{A\cup B}^{\rm early} = 2S_0 + \frac{4\pi c t}{3\beta} \ ,
\ee
where
\be \label{eq:S0}
2S_0 = S_{A \cup B}(t=0) = \frac{2c}{3} \log\frac{\beta}{2 \pi \epsilon}\, ,
\ee
is a constant and we have made explicit the dependence on a UV-cutoff $\epsilon$. 

\subsubsection*{Late times}
Now consider
\be
t > \frac D 2 + L \ .
\ee
In this regime, the $\beta \to 0$ limit sends
\be\label{etazero}
\eta_{ij} \to 0 \ ,
\ee
with
\be
\eta_{12} \sim \eta_{34} \ll \eta_{23} \ll \eta_{24} \sim \eta_{13} \ll \eta_{14} \ .
\ee
In this limit, $w_2 \to w_1$ and $\bw_2 \to \bw_1$ faster than any other limit, so we can apply the OPE $(w_2, \bw_2) \to (w_1, \bw_1)$ in the manner of Eq.~\eqref{limita}. This leaves the operators at points 3 and 4, and since $\eta_{34} \to 0$ we can apply the OPE $(w_3, \bw_3) \to (w_4, \bw_4)$.  
Therefore
\bea\label{lateuhp}
I_n^{\text{UHP}} &\sim& \left[ (w_1-w_2)(w_3-w_4)(\bw_1-\bw_2)(\bw_3-\bw_4)\right]^{-2h_n}\\
&\sim& \exp\left[-\frac{8\pi}{\beta}h_n (D+3L)\right] \notag \ .
\eea
This gives the entanglement entropy
\be\label{slate}
S_{A\cup B}^{\rm late} = 2S_0 + \frac{2\pi c L}{3\beta} \ .
\ee
The second term is the thermal value, $(2L) \times s_{\text{thermal}}$, where $s_{\text{thermal}}$ 
is the thermal entropy density at inverse temperature $\beta$.

\subsubsection*{Intermediate times, before the dip}
Now consider
\be
\frac{L}{2} < t < \frac{D}{2} \ .
\ee
The upper limit restricts to the time before the possible dip. In this case,
\be
\eta_{12} \to 0 , \quad \eta_{13} \to 1 , \quad \eta_{14} \to 1 , \quad \eta_{23} \to 1 , \quad \eta_{34} \to 0 \ .
\ee
Once again we can use the OPEs $(w_2, \bw_2) \to (w_1, \bw_1)$ and then $(w_4, \bw_4) \to (w_3, \bw_3)$, so the answer is the same as at late times:
\be
S_{A \cup B}^{\rm intermed} = 2S_0 + \frac{2\pi c L}{3\beta} \ .
\ee


\subsection{Light cone singularities and the dip}\label{ss:lcdip}
\begin{figure}
\centering
\includegraphics{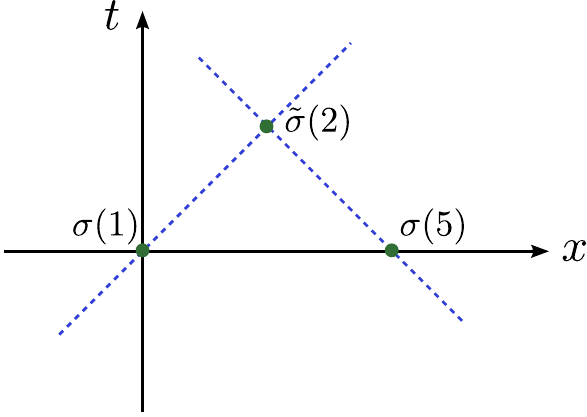}
\caption{Light cone limit $x \to 0$, $\bx \to 1$.\label{fig:diamond}}
\end{figure}

Finally we turn to the dip regime,
\be\label{dipregime}
\frac{D}{2} < t < \frac{D}{2} + L \ .
\ee
In this regime, for $t < \frac{D}{2} + \frac{L}{2}$, the limit $\beta\to 0$ corresponds to
\be\label{dipa}
\eta_{12} \to 0 , \quad \eta_{13} \to 1 , \quad \eta_{14} \to 1 , \quad \eta_{23}\to 0 , \quad \eta_{24} \to 1 , \quad \eta_{34} \to 0 \ ,
\ee
and for $t > \frac{D}{2} + \frac{L}{2}$,\footnote{\eqref{dipb} illustrates why we need to be careful about the order of limits --- it is a different limit than \eqref{etazero}, but would seem to be identical if we only looked at the five independent cross ratios $(\eta_{12}, \eta_{13}, \eta_{23}, \eta_{24}, \eta_{34})$ and ignored the order of limits.}
\be\label{dipb}
\eta_{12} \to 0 , \quad \eta_{13} \to 0 , \quad \eta_{14} \to 1 , \quad \eta_{23}\to 0 , \quad \eta_{24} \to 0 , \quad \eta_{34} \to 0 \ .
\ee
Now we come to a key point:  the configurations in eqs.~\eqref{dipa} or \eqref{dipb} do not correspond to any combination of
the OPE limits \eqref{limita} and \eqref{limitb}. Instead, they correspond to the limits
\bea\label{wwlim}
&&w_1 \to w_2 , \quad w_3 \to w_8 , \quad w_4 \to w_7 , \quad w_5 \to w_6\\
&& \bw_1 \to \bw_6 , \quad \bw_2 \to \bw_5 , \quad \bw_3 \to \bw_4 , \quad \bw_7 \to \bw_8 \notag \ .
\eea
This is not possible with the OPE, because the $w_i$ are in a different channel than the $\bw_i$.  These limits are therefore intrinsically Lorentzian. They correspond to an operator hitting the light cones of two other operators, simultaneously.  To illustrate this, consider the cross ratio
\be\label{xcross}
x = \frac{(w_1 - w_2)(w_5 - w_6)}{(w_1 - w_5)(w_2 - w_6)} \ .
\ee
In the regime \eqref{dipregime}, 
\be\label{mixlimit}
x \to 0 \ , \qquad \bx \to 1 \ .
\ee
If we map to $w_1 = \bw_1 = 0 , w_5 = \bw_5 = 1, w_6 = \bw_6 = \infty$, this corresponds to the limit where $(w_2, \bw_2)$ approaches the tip of the causal diamond bounded by the light cones of operators 1 and 5, as in figure \ref{fig:diamond}. 

In general, the OPE does not fix the behavior of a correlator in this type of limit, and in particular it does not imply that there is a singularity $[x(1-\bx)]^{-2h_n}$.  On the other hand, in rational CFT, these singularities \textit{do} exist.  Much of the rest of the paper will be devoted to studying these singularities in detail, and understanding the difference in the non-rational case.\footnote{ In \cite{Calabrese:2005in,Calabrese:2009qy} it was assumed that the singularity $[x(1-\bx)]^{-2h_n}$ always exists. In the notation of \cite{Calabrese:2009qy}, this assumption is implicit in the claim that ${\cal F}_{n,N}$ can be set to one in the limit $\beta \to 0$. This is the technical origin of the difference between our conclusions and previous analyses.}

To see when we do get a strong dip in the entanglement entropy, suppose that we can treat $w$ and $\bw$ independently, so that we can choose different channels for left and right movers.  Then the limit \eqref{wwlim} would produce a term
\bea\label{lcs}
I_n^{\text{UHP}} &\sim& [(w_1 - w_2)(w_3 - w_8)(w_4 - w_7)(w_5 - w_6) \\
& & \qquad \times (\bw_1 - \bw_6)(\bw_2 - \bw_5)(\bw_3 - \bw_4)(\bw_7 - \bw_8)]^{-h_n}\notag \ .
\eea
This dominates over the other channels in the regime \eqref{dipregime}, and leads to an entanglement entropy
\be\label{dipee}
S_{A\cup B}^{\rm dip}(t) = 2S_0 +  \frac{\pi c}{3\beta}\left\{ \begin{array}{cc}
D+2L-2t & \qquad \frac{D}{2} < t< \frac{D}{2} + \frac{L}{2}\\
 2t-D & \qquad \frac{D}{2} + \frac{L}{2} < t < \frac{D}{2} + L
 \end{array}\right. \,.
\ee
This reproduces the dip of \cite{Calabrese:2005in,Calabrese:2007rg,Calabrese:2009qy}, plotted in figure \ref{fig:dip}. 

To summarize: \textit{(i)} this dip exists if and only if the 4-point function of twist operators in the UHP has the light cone singularity \eqref{lcs}, and \textit{(ii)} this singularity is not required by the OPE (though we will see that it is required by crossing symmetry in rational CFT). 

Note that the additional light cone singularity we are referring to is not simply the singularity where an operator in the UHP hits the image point of another operator.  This type of singularity, which is also not required by OPE, is not sufficient to produce the quasiparticle dip.


\section{Light cone singularities in the R\'enyi entropy}\label{sec:renyi}

\subsection{Thermal double model}
The method of images determines the conformal transformation properties of a correlator on the UHP, but it cannot be used to actually compute the correlator.  In other words, the 4-point function of twist operators on the UHP is not fixed by the 8-point function with image points of twist operators on the full plane. In particular it depends on the specific boundary condition.  To avoid this complication, we can use a slightly different model for the quench introduced in \cite{Hartman:2013qma}: simply replace the image points by actual twist operators, \ie compute the 8-point function
\be\label{inplane}\small
\langle \sigma(w_1, \bw_1) \tsigma(w_2, \bw_2) \sigma(w_3, \bw_3) \tsigma(w_4, \bw_4) \sigma(w_5, \bw_5) \tsigma(w_6, \bw_6) \sigma(w_7, \bw_7) \tsigma(w_8,\bw_8)\rangle_{\text{plane}} \, .
\ee

\begin{figure}
\centering
\includegraphics{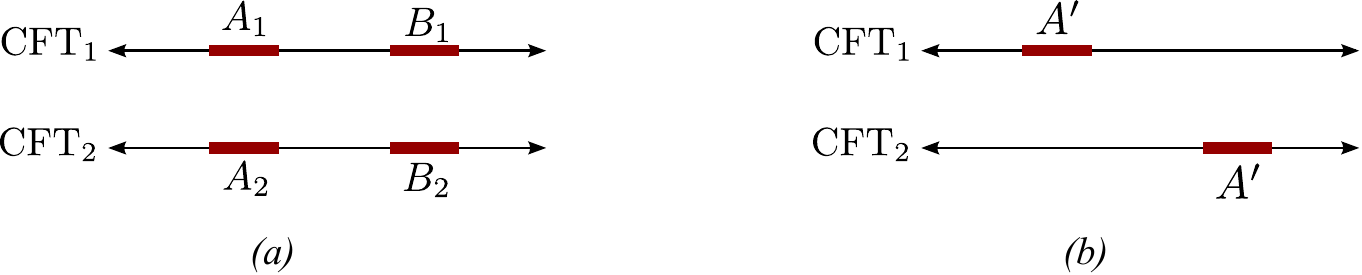}
\caption{\textit{(a)} Entanglement regions in the thermal double model, corresponding to $S_{A\cup B}$ in the original CFT.  \textit{(b)} Simplified setup, with an interval in the CFT and an offset interval in the thermal double.  This setup exhibits the same features, including long-term memory effects in the quasiparticle picture that are not necessarily present in CFT. \label{fig:doublemodel}}
\end{figure}

Physically, this calculates the R\'enyi entropy in a doubled CFT.  The doubled CFT consists of two decoupled copies of the original theory, CFT$_1 \times$ CFT$_2$, in the thermofield double entangled state
\be
|\text{TFD}\rangle = \sum_{n}e^{-\beta H/2}|n\rangle_1 |n\rangle_2 \ .
\ee
Each CFT is labeled by its own coordinates, $(t_1, x_1)$ or $(t_2, x_2)$.  The regions now include a piece in each system:
\be
A = A_1 \cup A_2 , \quad B = B_1 \cup B_2
\ee
where $A_{i=1,2}$ is the interval $x_i \in [0, L]$ in system $i$, and $B_i$ is the interval $x_i \in [D+L, D+2L]$.  See figure \ref{fig:doublemodel}$a$. The CFTs are evolved in time under the total Hamiltonian $H = H_1 + H_2$. 

The R\'enyi entropies are evaluated at $t_1 = t_2 = t$ and have a non-trivial time dependence in the 
$|\text{TFD}\rangle$ state.
Similarly to what we discussed in sec.~\ref{sec:bdrystate} for the boundary state model, 
they can be obtained from analytic continuation of the twist correlator for 
Euclidean insertions on a thermal cylinder of periodicity $\beta$. The twist insertions 
are at the endpoints of $A_i, B_i$ with a shift $i \beta/2$ for operators in two different copies of the CFT. The 8-point function on the plane \eqref{inplane} is related via conformal mapping to this 8-point function on the cylinder.

This thermal double model was designed partly to reproduce the physics of the Calabrese-Cardy boundary state model for a single interval.  For multiple intervals, it is not clear that these models are equivalent.  If anything, it appears that \eqref{inplane} may have more singularities than the corresponding quantity $I_n^{\text{UHP}}$ in the boundary state model, since we do not know of any argument that the boundary state correlator has singularities when insertions in the UHP hit image points of different operators in the lower half plane (for example $w_1 \to \bw_3$). Therefore we suspect, but have not proved, that the Calabrese-Cardy model cannot have any singularities beyond those in the thermal double. In this section we will determine what distinguishes the CFTs with and without a large dip in the R\'enyi entropies, in the thermal double model. With the (in our view conservative) additional assumption that boundary correlators do not have any additional singularities beyond those present in the 8-point function, the same conclusions apply to the boundary state model.
 

\subsection{Offset intervals in the thermal double}\label{sec:offset}

We have argued that entanglement memory comes from the light cone singularity of the cross-ratio $x$ introduced in \eqref{xcross}. To isolate this physics, we will study the simpler correlator
\be\label{rcor}
I_n^{\text{plane}} \equiv   \langle \sigma(w_1, \bw_1) \tsigma(w_2, \bw_2) \sigma(w_5, \bw_5) \tsigma(w_6, \bw_6)\rangle_{\text{plane}}  \ ,
\ee
with the same points as defined above,
\begin{align}
w_1 &= e^{\frac{2\pi}{\beta}(-t+i\beta/4)} ,  &  \bw_1 &= e^{\frac{2\pi}{\beta}(t-i\beta/4)}\\
w_2 &= e^{\frac{2\pi}{\beta}(L-t+i\beta/4)} , & \bw_2 &= e^{\frac{2\pi}{\beta}(L + t - i\beta/4)}\notag\\
w_5 &= e^{\frac{2\pi}{\beta}(D+2L+t-i\beta/4)} , & \bw_5 &= e^{\frac{2\pi}{\beta}(D+2L-t+i\beta/4)}\notag\\
w_6 &= e^{\frac{2\pi}{\beta}(D+L+t-i\beta/4)} , & \bw_6 &= e^{\frac{2\pi}{\beta}(D+L-t+i\beta/4)}\notag \ .
\end{align}
If there is a dip in the entanglement computed from the 8-point function, then it must also appear in this 4-point function. So, if we can rule out quasiparticle-like singularities of this 4-point function, then the quasiparticle picture is ruled out for the 8-point function as well.

\begin{figure}
\centering
\includegraphics{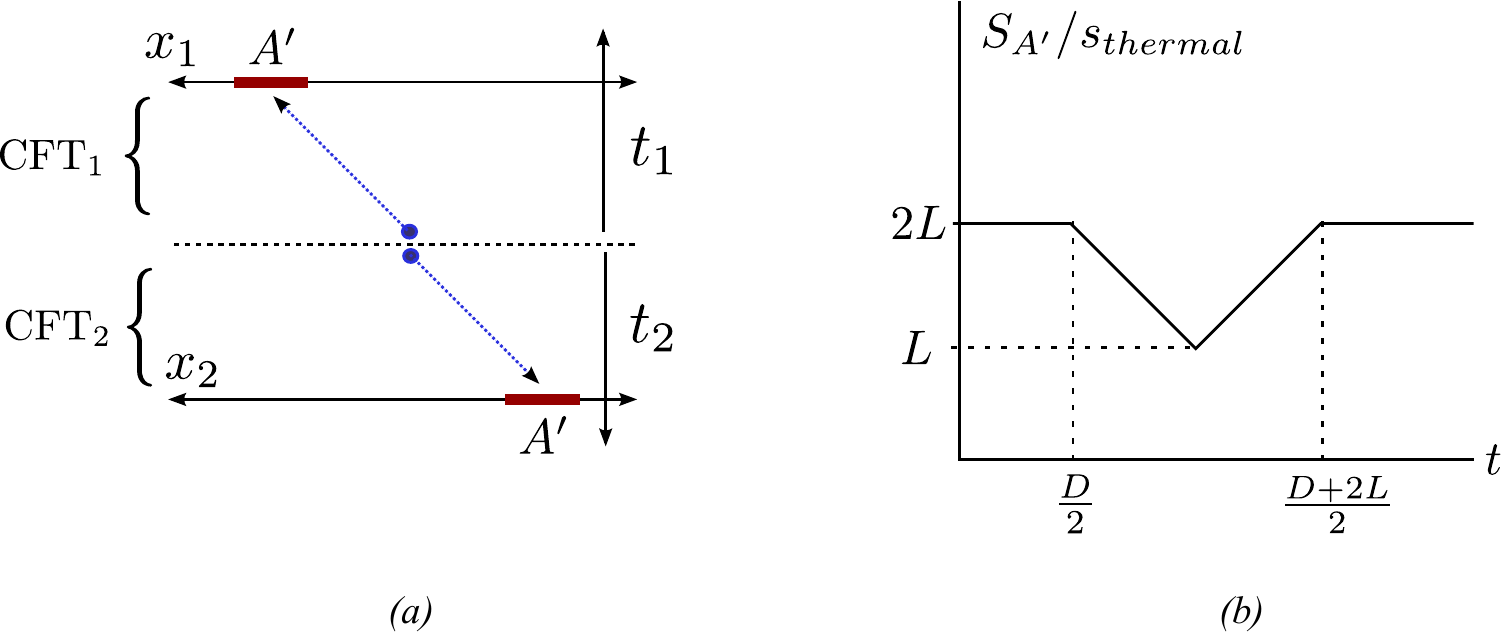}
\caption{Quasiparticle picture for the offset-intervals entanglement in the thermal double model. \textit{(a)} Entangled quasiparticle pairs (blue dots) in systems 1 and 2 are separated by $\Delta x \sim \beta$ at $t = 0$ (dashed horizontal line). Under time evolution, the particles move in opposite directions. Eventually both particles enter region $A'$. \textit{(b)} Quasiparticle prediction for the entanglement entropy $S_{A'}-2S_0$ of region $A'$.  When the entangled pair enters region $A'$, the entanglement entropy decreases. \label{fig:thermalqp}}
\end{figure}

The correlator \eqref{rcor} has a natural physical interpretation in the thermal double theory:  It computes the R\'enyi entropy of a region $A'$ consisting of an interval in CFT$_1$, and an offset interval in the thermal double CFT$_2$, as in figure \ref{fig:doublemodel}$b$:
\be
A': \{ x_1 \in [0, L]\} \cup \{ x_2 \in [D+L, D+2L]\} \ .
\ee
Namely
\be \label{eq:cylinder}
\Tr \rho_{A'}^n  \sim I^{\text{cylinder}}_n = \left(2\pi \over \beta\right)^{8h_n} |w_1 w_2 w_5 w_6|^{2h_n}I_n^{\text{plane}}  \,.
\ee
The quasiparticle picture predicts a memory effect in this situation.  In the thermal double, the initial state in the quasiparticle picture consists of entangled pairs, where one member of each pair is in CFT$_1$ and the other is in CFT$_2$, with both particles at the same $x$-position.  All of these pairs contribute to the entanglement entropy at $t=0$, so the initial entanglement entropy is $(2L) \times s_{\text{thermal}}(\beta)$ (plus the usual divergent term). Under time evolution, the particle in CFT$_1$ moves one direction, and the particle in CFT$_2$ moves the other direction.  In the range
\be\label{diprange}
\frac{D}{2} < t < \frac{D}{2} + L
\ee
we therefore find particles in CFT$_1$ with $x_1 \in [0, L]$ that are entangled with particles in CFT$_2$ with $x_2 \in [D+L, D+2L]$. This is illustrated in figure \ref{fig:thermalqp}$a$. Both of these particles are in region $A'$, so the entanglement entropy decreases.  Thus the quasiparticle prediction for the entanglement entropy, corresponding to the $n\to 1$ limit of \eqref{rcor}, is as shown in figure \ref{fig:thermalqp}$b$.   The early-time prediction is exactly $S_{A\cup B}^{\text{late, intermed}}$ computed for the boundary state in \eqref{slate}, and the dip prediction is the same as $S_{A\cup B}^{\text{dip}}$ in \eqref{dipee}. Our goal in this section is to understand when this quasiparticle prediction is correct and when it is incorrect.

The correlator \eqref{rcor} is (for $t,L,D\gg\beta$)
\be\label{tostrip}
I_n^{\text{plane}}  =\exp\left[-\frac{8\pi}{\beta}h_n(D+3L)\right](x \bx)^{2h_n}  G_n(x,\bx) \ ,
\ee
where
\be\label{gxx}
G_n(x,\bx) \equiv \langle \sigma(0) \tsigma(x,\bx) \sigma(1) \tsigma(\infty)\rangle \ .
\ee
The cross-ratio $x$, defined in \eqref{xcross}, is
\bea
x &\approx& \exp\left[-\frac{2\pi}{\beta}(D+2t)\right]  \approx 0\\
\bx &\approx& \exp\left[\frac{2\pi }{\beta}( D+2L+t - \max(D+2L-t,t) - \max(D, 2t) ) \right] \ .\notag
\eea
The OPE implies that this correlator has the singularities
\be\label{zerosing}
G_n \sim (x \bx)^{-2h_n} \qquad (\mbox{as\ } x,\bx \to 0)
\ee
and 
\be\label{onesing}
G_n \sim [(1-x)(1-\bx)]^{-2h_n} \qquad (\mbox{as\ } x, \bx \to 1) \ .
\ee
Away from the dip, \ie outside the range \eqref{diprange}, $\bx \approx 0$ so the correlator is given by \eqref{zerosing}.  This leads to the entropy  
\be
S_{A'}^{\text{early, late}} = 2S_0 + \frac{2\pi c L}{3\beta}  \ ,
\ee
as expected from the quasiparticle picture.

In the range \eqref{diprange}, $x\to 0 $ and $\bx \to 1$, with
\be
1 - \bx \approx \exp\left[-\frac{2\pi}{\beta}\min( D+2L-2t, 2t-D) \right] \ .
\ee
so that
\be
\frac{x}{1 - \bx} \to 0 \ .
\ee  
The OPE does not require any particular behavior in this limit.  But if we assume that $G_n(x,\bx) \sim [x(1-\bx)]^{-2h_n}$ in this limit, then the resulting entanglement entropy is equal to \eqref{dipee}.
This would agree with the quasiparticle picture.  Therefore in this simple setup, the validity of the quasiparticle model is precisely the question of whether the correlator \eqref{gxx} has a singularity
\be\label{gfsing}
G_n(x,\bx) \stackrel{?}{\sim} [x(1-\bx)]^{-2h_n} \qquad (\mbox{as\ } x \to 0, \bar x \to 1) \,.
\ee


\subsection{Light cone features in the second R\'enyi}\label{sec:2ndRenyi}
Our conclusion will be that the singularity \eqref{gfsing} exists if and only if the theory is current-dominated, as explained in the introduction. It is simplest to make this argument for the second R\'enyi entropy,
\be
G_2(x,\bx)= \langle \sigma_2(0) \tsigma_2(x,\bx) \sigma_2(1) \tsigma_2(\infty)\rangle \ ,
\ee
where the subscript indicates replica number $n=2$. The two-sheeted manifold that defines these double-twist operators is in fact a torus.  It follows that this correlation function is fixed by the torus partition function, and the mapping derived in \cite{Lunin:2000yv,Headrick:2010zt} is
\be\label{gtoz}
G_2(x,\bx) = \(2^{16} x \bx (1-x)(1-\bx)\)^{-c/24} Z(\tau, \btau) \ .
\ee
The partition function,
\be
Z(\tau, \btau) = \Tr \exp\left[2\pi i \tau\(L_0 - \frac{c}{24}\) - 2\pi i \btau \(\bar{L}_0 - \frac{c}{24}\)\right] \ ,
\ee
is evaluated with the torus modulus related to the twist operator insertions by
\be\label{xtau}
x = \frac{\theta_2(\tau)^4}{\theta_3(\tau)^4}\,  , \qquad  \tau = i \frac{K(1-x)}{K(x)} \ ,
\ee
where $K(x) = \frac{\pi}{2}\, _2F_1(\half,\half,1,x^2)$, and similarly for $\btau(\bx)$. For small $x$, the relation is
\be\label{qtau}
x = 16 \sqrt{q} + O(q) , \qquad q\equiv e^{2\pi i \tau} \ ,
\ee
and for $x$ near 1, 
\be
1-x = 16 \sqrt{q'} + O(q') , \qquad q' \equiv e^{-2\pi i/\tau} \ .
\ee
The modular transformation of the torus $\tau \to -1/\tau$ is related to crossing symmetry of the 4-point function, $x \to 1-x$.

This maps the problem of memory effects in the 2nd R\'enyi to a question about the spectrum.  Including the prefactors from \eqref{gtoz} and \eqref{tostrip}, and the conformal factor in  \eqref{eq:cylinder} from mapping the cylinder to the plane, the 2nd R\'enyi is
\be\label{stgen}
S_2 =  \frac{\pi c L}{2\beta} - \log \left[2^{-2c/3}(x \bx)^{c/12}[(1-x)(1-\bx)]^{-c/24} Z(\tau(x), \btau(\bx))\right]  + S_2^0\ ,
\ee
where
\be
 S_2^0 = \frac c 2 \log \frac{\beta}{2 \pi \epsilon}
 \ee
 is a constant independent of $t,L,D$.

First, let's reproduce the OPE singularities \eqref{zerosing} and \eqref{onesing}.  The limit $x, \bx \to 0$ is the zero temperature limit of the partition function $\tau \to i \infty, \btau \to - i \infty$, so only the vacuum contributes to the trace:
\be\label{eq:ZlowT}
Z(\tau, \btau) \sim q^{-c/24} \bq ^{-c/24} \sim 2^{2c/3} x^{-c/12}\bx^{-c/12} \ .
\ee
Including the prefactor in \eqref{gtoz}, 
\be
G_2 \sim (x \bx)^{-c/8} = (x \bx)^{-2h_2} \ .
\ee
The other OPE limit, $x, \bx \to 1$, is a high temperature limit.  In this limit, we use modular invariance of the torus partition function, projecting the trace onto the vacuum in the dual channel 
\be
Z(\tau,\btau) \sim (q')^{-c/24}(\bq')^{-c/24}
\ee
which leads to
\be
G_2\sim [(1-x)(1-\bx)]^{-2h_2} \ ,
\ee
as expected.

Now consider the mixed limit $x \ll 1-\bx \ll 1$.  This is a mixed-temperature limit, where we send 
\be
\tau \to i \infty, \qquad \btau \to i 0^- \ .
\ee
This is low temperature for left movers, and high temperature for right movers. We first take the limit $x \to 0$, so that the partition function gets contributions only from left-moving ground states, \ie conserved currents,
\be\label{zdd}
Z(\tau, \btau) \sim q^{-c/24} \sum_{{\rm currents}}\bq^{\bL_0 - c/24} \ .
\ee
The sum is over states with conformal weights $(0, \bL_0)$. The limit that is actually relevant to the 2nd R\'enyi simultaneously takes $x\to 0$ and $\bx \to 1$ in a particular ratio, so \eqref{zdd} might not contain the dominant terms in that limit.  However our goal is only to check whether the quasiparticle singularity is present, and \eqref{zdd} is sufficient for this purpose.

Now in the limit $\bx \to 1$, the sum is dominated by the states with very large $\bL_0$.  The density of states in CFT, $\Omega(L_0, \bL_0)$, is fixed asymptotically by the Cardy formula \cite{Cardy:1986ie},
\be
\Omega(L_0, \bL_0) \approx \exp\left( 2\pi \sqrt{\frac{c}{6}L_0} + 2\pi\sqrt{\frac{c}{6}\bL_0} \right) \qquad (\mbox{as\ } L_0, \bL_0 \to \infty) \ ,
\ee
where $c$ is the central charge.  This formula applies only when both $L_0$ and $\bL_0$ are large, so it does not fix the behavior of \eqref{zdd}.  Instead, define the density of currents
\be
\Omega_{\rm currents}(\bL_0) = \Omega(L_0 = 0, \bL_0) \ .
\ee
The singularity in \eqref{zdd} will be fixed by the asymptotic growth of $\Omega_{\rm currents}$.  Let us parameterize this growth by a number $c_{\rm currents}$, defined such that
\be
\Omega_{\rm currents}(\bL_0) \approx \exp\left( 2\pi \sqrt{\frac{c_{\rm currents}}{6}\bL_0} \right) \qquad \mbox{as} \qquad \bL_0 \to \infty \ .
\ee
 Clearly $c \geq c_{\rm currents}$. To see that we can always parameterize the asymptotic growth of $\Omega_{\rm currents}$ in this way, first note that the growth must be at least this fast, since the stress tensor guarantees $c_{\rm currents}\geq \frac{1}{2}$.\footnote{Without null states, the stress tensor descendants of the vacuum grow asymptotically as $\Omega \sim \exp (\pi \sqrt{2 \bar L_0/3} )$, corresponding to $c_{\rm currents} = 1$. With null states this number can decrease, with the lower bound $c_{\rm currents} = \half$ set by the Ising model since this is the unitary minimal model of lowest central charge.} Also, the exponent cannot have a larger power of $\bL_0$, since a singularity stronger than the quasiparticle singularity would not be consistent with modular invariance.

 $c_{\rm currents}$ can be interpreted as an effective central charge for the conserved right-moving current sector of the theory.  As $\bq \to 1$, the sum \eqref{zdd} can be evaluated by a saddlepoint approximation, with the usual Cardy result 
\be\label{zgen}
Z(\tau, \btau) \sim q^{-c/24} (\bq')^{-c_{\rm currents}/24}  \sim 2^{(c+c_{\rm currents})/3} x^{-c/12}(1-\bx)^{-c_{\rm currents}/12}\ .
\ee
This formula is one of our main results: it implies that the quasiparticle picture for the second R\'enyi applies if and only if $c_{\rm currents} = c$.  That is, the quasiparticle picture holds only in theories which are current dominated, meaning the asymptotic number of conserved currents is approximately equal to the total number of states.  This is true in rational CFT (see below), but not true in general.  In particular, a theory with no conserved currents besides the stress tensor has 
\be
c_{\rm currents} \leq 1 \ .
\ee
Thus, in a theory with $c>1$ and no extended chiral symmetry, the quasiparticle picture does not produce the correct 2nd R\'enyi entropy.\footnote{In the $W_N$ minimal models, which generalize the Virasoro minimal models to a $W_N$ chiral algebra, $c_{\rm currents} < N-1$. In a theory with chiral algebra $W_N$, but $c > N-1$, there are no null states in the vacuum representation and this leads to $c_{\rm currents} = N-1$.}

In a non-rational theory with $c_{\rm currents}<c$, the 2nd R\'enyi entropy is not universal, since representations other than the vacuum may dominate the partition function.  The contribution from the vacuum representation, obtained by inserting  \eqref{zgen} into \eqref{stgen}, sets an upper bound during the dip regime, and the quasiparticle result sets a lower bound:
\begin{align} \label{eq:2ndRenyidip}
\frac{\pi c L}{2\beta} - \frac{\pi c}{4\beta} & \min(D+2L-2t, 2t - D) +S_2^0 \\
&< S_2^{\rm dip} \leq  \frac{\pi c L}{2\beta} - \frac{(c+2c_{\rm currents})\pi}{12\beta} \min(D+2L-2t, 2t - D) +S_2^0 \ .\notag
\end{align}
For $c_{\rm currents} = c$, the dip in $S_2$ is as large as possible; for $c_{\rm currents} \ll c$ there is still a dip but its magnitude is smaller by a factor of at most 3.

\subsubsection*{More comments on rational vs general CFTs}
In a rational CFT  such as the $c<1$ minimal models, the partition function is a finite sum of characters:\footnote{We assume the diagonal modular invariant but the conclusions are similar for other invariants.}
\be
Z(\tau, \btau) = \sum_{R}\chi_R(q) \chi_R(\bq) \ .
\ee
These characters transform under $S: \tau \to -1/\tau$ by the action of the modular $S$-matrix,
\be
\chi_R(q) = \sum_P S_{RP}\chi_P(q') \ .
\ee
This is again a finite sum. It follows that the mixed-temperature limit of the partition function relevant to the R\'enyi with $x \ll 1-\bx \ll 1$ is
\be
\label{eq:Z_rat}
Z(\tau,\btau) \sim S_{00} q^{-c/24} (\bq')^{-c/24} \ .
\ee
Comparing to \eqref{zgen}, we see $c = c_{\rm currents}$ in rational CFT.

In a $c>1$ CFT where the vacuum representation has only Virasoro descendants, the manner in which modular invariance is enforced is quite different.  The $S$-transformation of the vacuum character is an integral\cite{Zamolodchikov:2001ah}\footnote{
$S(h,0) =  \sqrt{2\over p_h }\sinh(2\pi b p_h) \sinh(2\pi b^{-1} p_h)$ where $h=p_h^2 + \frac{c-1}{24}$ and $c = 1 + 6(b + b^{-1})^2$.}
\be\label{vactrans}
\chi_0^{\text{Vir}}(\bq) =  \int_{(c-1)/24}^\infty dh S(h,0) \chi_h^{\text{Vir}}(\bq') \ .
\ee
This is clearly not a sum over characters in the dual channel, and does not include a contribution from the vacuum. As $\bq \to 1$,
\be
\chi_0^{\text{Vir}}(\bq) \sim (\bq')^{-1/24}
\ee
so $c_{\rm currents} = 1$. Modular invariance still holds, but in a non-rational CFT, the vacuum singularity as $\bar \tau \to i 0^-$ does not come from any individual character in the original trace. Instead it comes from the asymptotics of the infinite sum.


\subsection{Conformal block expansion} \label{ss:cbe}
We now return to the general R\'enyi index $n$. In this case we can make a similar argument using conformal blocks instead of characters, but the singularity in the dual channel is poorly understood.

The twist correlator may be expanded in conformal blocks in any channel. In the $s$-channel $x \to 0$,
\be\label{exps}
G_n(x,\bx) = \sum_p a_p F_p(cn, h_n; x)F_p(cn, h_n; \bx) \ ,
\ee
where $a_p$ is a constant related to the OPE coefficients, and $F_p(c',h;x)$ is the conformal block with internal weight $h_p$, external weights $h$, and central charge $c' = cn$ (since this is a correlator in the replica theory). The conformal blocks are, by definition, fixed entirely by the chiral algebra. In the limit $x\ll 1 - \bx \ll 1$, the dominant contribution is not necessarily the vacuum. The contribution from the vacuum sets a lower bound,
\be\label{nren}
G_n(x, \bx) \gtrsim F_0(cn, h_n; x)F_0(cn, h_n; \bx)  \approx x^{-2h_n}F_0(cn, h_n; \bx) \ .
\ee
This contribution is universal, in the sense that it does not depend on the full details of the theory: it depends only on the conformal block $F_0(cn, h_n; \bx)$ in the limit $\bx \to 1$. It is therefore fixed by the right-moving chiral algebra. Of course, if the two intervals are offset in the opposite direction, then left-moving currents contribute instead. In the original quench model, with two intervals in a pure state of a single CFT, both types of currents contribute.

The conformal block expansion \eqref{exps} is in the orbifold, CFT$^n/\mathbb{Z}_n$, so $F_0$ in \eqref{nren} is the vacuum block in the orbifold.  The chiral algebra of the orbifold is fixed by the chiral algebra of original CFT, but is not identical (see for example \cite{Headrick:2010zt}). This distinction is discussed further below.

\subsubsection*{Rational CFT has the quasiparticle dip}
In a rational CFT,  we can expand any individual conformal block as a finite sum over primaries in the dual channel:
\be
F_p(\bx) = \sum_{q} b_{pq} F_q(1-\bx) \ ,
\ee
(with some of the arguments of the conformal block suppressed).
Every term contributes on the right hand side.  Thus
\be\label{bbz}
F_0(\bx) \sim b_{00} (1-\bx)^{-2h_n} \quad \mbox{as} \quad \bx \to 1 \ .
\ee
This is the quasiparticle singularity \eqref{gfsing}.\footnote{The coefficient $b_{00}$ in \eqref{bbz} is related to the quantum dimension of the twist operator in the orbifold theory.  It does not affect the answer in the scaling limit $t,L,D  \gg \beta$, but may have interesting implications at intermediate times; quantum dimensions have appeared several times before in entanglement calculations \cite{Kitaev:2005dm,Levin:2006zz,He:2014mwa}.}  Thus in the scaling limit $t, L,D\gg \xi$, the quasiparticle picture does apply to rational CFTs.

One way to understand the origin of the singularity in $F_0(\bx)$ as $\bx \to 1$ in rational CFT is from crossing symmetry,
\be\label{crossing}
G_n(x, \bx) = G_n(1-x, 1-\bx) \ .
\ee
In a rational CFT, each side of this equation is a finite sum.  Therefore, in order to reproduce the identity singularity on the rhs, individual terms on the lhs must go as $(1-\bx)^{-2h_n}$.


\subsubsection*{General CFT}

In a general CFT, the conformal block expansion is infinite.  In this case the crossing equation \eqref{crossing} does not imply anything about the singularities of individual terms in the $s$-channel as $\bx \to 1$, since these singularities can come from the infinite sum rather than from any particular term.  This is exactly what occurs in higher spacetime dimensions, where there is no such thing as rational CFT --- individual conformal blocks cannot reproduce the identity singularity in the other channel, so it is instead reproduced by the asymptotics of the infinite sum (see for example \cite{Pappadopulo:2012jk}). 

Suppose that the chiral algebra of the CFT is just the Virasoro algebra.  That is, the only states in the CFT of dimension $(0, \bL_0)$ are the Virasoro descendants of the vacuum.  The vacuum contribution to the twist correlator for $x \ll 1-\bx \ll 1$ is then
\be
G_n(x, \bx) \sim x^{-2h_n}F_0^{\text{Vir}^n/\mathbb{Z}_n}(cn, h_n; \bx) \ ,
\ee
where $F_0^{\text{Vir}^n/\mathbb{Z}_n}$ is the vacuum conformal block associated to the chiral algebra $\text{Vir}^n/\mathbb{Z}_n$.  This conformal block is not just the Virasoro block, though it is fixed entirely by the Virasoro algebra. Denoting the Virasoro modes in copy $k$ of the CFT by $\bL_n^{(k)}$, the block $F_0^{\text{Vir}^n/\mathbb{Z}_n}$ is defined to include the contributions from all $\mathbb{Z}_n$-symmetric combinations of the states
\be
\bL_{-n_1}^{(k_1)} \bL_{-n_2}^{(k_2)} \cdots \bL_{-n_r}^{(k_r)}|0\rangle \ .
\ee
 To proceed, we would need the behavior of this vacuum block in the limit $\bx \to 1$, but this is not known.  Indeed, even the behavior of $F_0^{\text{Vir}}$ in the limit $\bx \to 1$ appears to be unknown. However, there is generically no reason for it to reproduce the quasiparticle singularity \eqref{gfsing}. We therefore expect in this limit
\be\label{ffgo}
F_0^{\text{Vir}^n/\mathbb{Z}_n}(cn, h_n; \bx) \ll (1-\bx)^{-2h_n} \ .
\ee
Our results for the second R\'enyi entropy in section \ref{sec:2ndRenyi} confirmed \eqref{ffgo} for $n=2$.  We do not have a conclusive argument argument for $n >2$, but since the singularity is not required it is highly implausible that it would appear (for generic central charge --- of course it does appear in the minimal models). Crossing symmetry does not allow a singularity stronger than $(1-\bx)^{-2h_n}$, so this leaves \eqref{ffgo}. Other than the second R\'enyi, there is one other case we are aware of with $c>1$ where the Virasoro blocks can be computed exactly ($c=25, h=15/16$ \cite{zamof}), and the inequality analogous to \eqref{ffgo} holds in that case as well.\footnote{Explicitly:
\be
F_0\left(c = 25, h = \frac{15}{16}; \bx\right) = \left[ 16 \bq (\bx(1-\bx))^{7/8}\theta_3(\bq)^3\right]^{-1} \ , \qquad \bq \equiv \exp\left[-\pi \frac{K(1-\bx)}{K(\bx)}\right]  \ ,
\ee
where $\theta_3$ and $K$ are the standard elliptic functions. In the limit $\bx \to 1$, we have $\bq \sim e^{-\pi^2/\log(16/(1-\bx))}$ and $\theta_3(\bq) \sim \sqrt{\pi\over 1-\bq} $, so, ignoring constants and log corrections,
\be\label{specc}
F_0\left(c = 25, h = \frac{15}{16}; \bx\right) \sim (1-\bx)^{-7/8}  \qquad ({\mbox as\ } \bx \to 1) \ .
\ee
This is much weaker than the vacuum singularity $(1-\bx)^{-2h} = (1-\bx)^{-15/8}$.  }

Since we do not know the singularity of the orbifold conformal block as a function of $n$, we cannot determine the entanglement entropy, beyond the statement that it disagrees with the quasiparticle picture.  It would be very interesting to compute the universal vacuum contribution.


\subsection{$c=1$} \label{ss:c=1}

We have discussed the minimal models with $c<1$, and ``generic" CFTs with $c>1$ but no conserved currents other than the stress tensor.  
Now we turn to theories with $c = 1$ and show that the R\'enyi entropies 
of these theories agree with the quasiparticle picture.

The starting point for almost all known $c = 1$ CFTs is the compact free 
boson theory,\footnote{The theory in \cite{Runkel:2001ng} is an exception.}
defined by the Lagrangian 
\be
L = \frac{1}{2 \pi} \partial X \bar{\partial} X \ ,
\ee
with the compactification condition $X = X + 2\pi R$, i.e., 
the target space is a circle with radius $R$. This theory has 
$c = 1$ and infinitely many (Virasoro) conformal 
primaries with a discrete spectrum of conformal 
weights for finite $R$ \cite{Ginsparg:1988ui, DiFrancesco1997}.

We can analyze the light cone singularity of the second Reyni entropy of this 
theory explicitly, via the torus partition function \cite{DiFrancesco1997}
\beq
\label{eq:boson-pf}
	Z(R) = \frac{1}{\eta(\tau)\bar{\eta}(\bar{\tau})}  \sum_{e,m \in \mathbb{Z}} 
	q^{h_{e,m}} \bar{q}^{\bar{h}_{e,m}} \ ,
\eeq
where the conformal weights are given by 
\begin{align}
	h_{e,m} &= \frac12 \( \frac{e}{R} + \frac{mR}{2} \)^2 \, , \\
	\bar{h}_{e,m} &= \frac12 \( \frac{e}{R} - \frac{mR}{2} \)^2 \, ,
\end{align}
and $\eta$ denotes the Dedekind eta function.
When $R^2$ is rational, there are infinitely conformal primaries with a given $h_{e,m}$ 
or $\bar{h}_{e,m}$. For example, $h_{e,m} = 0$ whenever $2e/m = -R^2$.
Moreover, in this case the Virasoro algebra is extended to a 
larger chiral algebra, with respect to which there are finitely many 
primary operators, making the theory rational \cite{Dijkgraaf:1987vp, DiFrancesco1997}.
But when $R^2$ is irrational, there is no such degeneracy in the weights 
and the theory has no extended symmetry, so we investigate 
this case separately (these CFTs are examples of what \cite{Halpern:1995js} calls ``quasi-rational").

In the light cone limit $\tau \to i\infty, \bar{\tau} \to i0^-, q \to 0^+, \bar{q} \to 1^-$,
the partition function in Eq.~\eqref{eq:boson-pf} behaves as 
\beq
	Z(R) = \frac{1}{\eta(\tau)\bar{\eta}(\bar{\tau})}
	\left( 1 + q^{h_{\text{min}}} \bar{q}^{\bar{h}_{\text{min}}} + \cdots \right) \ ,
\eeq
where $h_{\text{min}}$ and $\bar{h}_{\text{min}}$ are the minimal non-zero holomorphic 
and anti-holomorphic conformal weights, respectively. 
The assumption that $R^2$ is irrational ensures that $h_{\text{min}}, \bh_{\text{min}} > 0$, \ie there are no additional conserved currents.
Therefore, all the terms besides $1$ approach $0$ in this limit, 
and the series is absolutely convergent,
so the behavior of the singularity is controlled by the $\eta$ functions. To analyze them we use the 
modular transformation property:
\beq
	\eta(-1/\tau) = \sqrt{-i\tau} \eta(\tau)\, , \quad \qquad \bar{\eta}(\bar{\tau}) =
	\sqrt{\frac{-i}{\bar{\tau}}} \bar{\eta}(\bar{\tau}') \ ,
\eeq
where $\bar{\tau}' \equiv -1/\bar{\tau}$.
In this limit $\bar{\tau}' \to -i\infty$ and so we have $\eta(\tau) \to q^{1/24}$ and 
$\bar{\eta}(\bar{\tau}') \to \bar{q}'^{1/24}$ and thus
\beq
	Z(R) 
	\sim  q^{-1/24} \bar{q}'^{-1/24} \ ,
\eeq
where the last expression gives the asymptotic behavior because the $q$ and $\bar{q}$ divergence 
is exponential and so dominates over the vanishing of $\bar{\tau}$.
We thus expect the dip in the second Reyni entropy after a quench, because we have 
the same asymptotic behavior as the rational CFT case, shown in eq.~\eqref{eq:Z_rat}.

The Reyni entropies of two intervals in the compact boson theory were computed in 
\cite{Calabrese:2009ez, Calabrese:2010he}, and generalized to arbitrary complex cross ratio in \cite{neg}. We will consider only the decompactified limit, $R\to \infty$. In this case the correlator of four $n$-twist operators is
\begin{align}
\label{bosonrenyi}
\langle \sigma(w_1, \bw_1) &\tsigma(w_2,\bw_2) \sigma(w_3,\bw_3) \tsigma(w_4, \bw_4)\rangle =\\
& \qquad  \left( w_{31}w_{42} \bw_{31} \bw_{42} \over w_{21}w_{43}w_{41}w_{32} \bw_{21}\bw_{43}\bw_{41}\bw_{32}\right)^{\frac{c}{12}(n-1/n)}{\cal F}_n(x,\bx) \ ,\notag
\end{align}
with $x = \frac{w_{21}w_{43}}{w_{31}w_{42}}$ and $w_{ij} = w_i-w_j$.
The explicit formula for ${\cal F}_n(x,\bx)$ is given in equation (142) of \cite{neg}, from which it is easy to check that the only power-law singularity in the light cone limit $w_2 \to w_1 , \bw_2 \to \bw_3$, which corresponds to $x \to 0, \bx \to 1$, comes from the prefactor in \eqref{bosonrenyi}.  This is the quasiparticle singularity.

The general picture, then, is that for a given chiral algebra, there is a critical central charge $c_{\rm currents}$. In many examples, including Virasoro symmetry, $W_N$ symmetry, and Kac-Moody symmetry (and perhaps in general), it is the same as the threshold for the theory to be rational.  If $c=c_{\rm currents}$, including irrational but `nearly rational' theories like the irrational boson where the central charge sits at the threshold, then the spectrum is current dominated and the entanglement entropy has quasiparticle behavior;  and if $c > c_{\rm currents}$ then the entanglement entropy disagrees with the quasiparticle picture.


\subsection{Entanglement is larger than quasiparticle at the dip midpoint} \label{ss:dipbound}

We have argued that in theories with $c>c_{\rm currents}$, the entanglement entropy of two intervals does not agree with the quasiparticle prediction.  There is one final step before we can conclude that memory effects are reduced: we need to argue that the dip is smaller in these theories, not larger. This seems obvious physically, but we will only give a proof limited to the second half of the dip regime: $\frac{D+L}{2} < t < \frac{D}{2}+L$.  This is enough to demonstrate the result at the midpoint of the quasiparticle dip:
\be\label{midi}
S_{A\cup B} \geq S_{A\cup B}^{\rm quasiparticle} \qquad \left(t = \frac{D+L}{2} \right) \ .
\ee
To demonstrate this, we apply strong subadditivity to the neighboring regions
\be
A: [0,L] \ , \qquad C: [L+\delta, L+D-\delta] , \qquad B: [L+D, 2L+D] \ ,
\ee
separated by a small distance, $\delta \ll \beta \ll t,L,D$.  The entanglement entropy of each individual region is given by the 
Calabrese-Cardy formula for a single interval of size $\ell$ after a global quench:
\be
S_\ell(t) = S_0+ \frac{\pi c}{3\beta} \min(2t, \ell) \ .
\ee
We can also apply their multiple-interval result to neighboring regions 
(since for separations of order $\delta$ contributions  from possible light cone singularities would be suppressed as $\delta/\beta$ ):
\be
S_{A\cup C}  = S_{C\cup B} = 2 S_{0} + \frac{\pi c}{3 \beta}\min(2t, L+D) \ ,
\ee
and
\be
S_{A\cup B\cup C} = 3S_{0} + \frac{\pi c}{3\beta}\min(2t, 2L+D) \ .
\ee
Now strong subadditivity, in the form
\be
S_{A\cup B}  \geq S_{A\cup C\cup B} + S_{B}-S_{C \cup B}
\ee
implies
\be
S_{A \cup B} \geq 2 S_{0} + \frac{\pi c}{3\beta}\left[\min(2t,2L+D) +\min(2t,L) - \min(2t,L+D)\right] \ .
\ee
In the range $\frac{D+L}{2} < t<\frac{D}{2}+L$, this becomes
\be
S_{A\cup B} \geq 2 S_{0} + \frac{\pi  c}{3\beta}(2t-D) = S_{A\cup B}^{\rm quasiparticle} \ .
\ee
Thus the quasiparticle prediction is a theoretical lower bound for the entanglement entropy during the second half of the dip (the rising edge of the dip in figure \ref{fig:dip}), in particular at the midpoint \eqref{midi}.

It is interesting to note that a similar bound for the first half of the dip regime $\frac{D}{2}<t<\frac{D+L}{2}$ could be obtained from the monogamy inequality 
\be
S_{A\cup B} \geq S_{A \cup C\cup B} + S_A + S_B + S_C -S_{C\cup A} - S_{C\cup B} \, ,
\ee
 which is believed to hold in holographic theories  \cite{Hayden:2011ag,Wall:2012uf}, but is not satisfied in general quantum systems.
 

\section{The large-$c$ limit}\label{s:largec}

We have argued that entanglement scrambles for $c$ above a critical value, but the actual value of the entanglement entropy in theories that scramble is not universal during the dip regime.  In this section, we will show that in the holographic $c\to \infty$ limit, all theories scramble \textit{maximally}. We will derive this universal limit using semiclassical gravity, and confirm it in conformal field theory (subject to some assumptions about the growth of OPE coefficients in these theories).  

Holographic CFTs have few chiral states, so $c_{\rm currents} \sim 1$, but many total states, so $c \gg 1$.  Therefore some of the CFT calculations in sections \ref{s:origin} and \ref{sec:renyi} must be revisited, since the large-$c$ limit may compete with the scaling limit $t,L,D \gg \beta$.  In section \ref{ss:stringy}, we address the interplay between these two limits (see also \cite{Caputa:2014vaa}) using the example of the 2nd R\'enyi in the D1-D5 CFT.


\subsection{Holographic entanglement entropy}\label{sec:HM}

The holographic dual of the thermofield double setup described in section~\ref{sec:renyi} has been studied in \cite{Hartman:2013qma}. It consists of the eternal BTZ black string with time taken to run forwards on both sides of the Penrose diagram, thus producing a time dependent excited state. For convenience, in this section we will set the inverse temperature $\beta$ of the black string to $2\pi$, and only reinstate $\beta$  in the final expression of the holographic entanglement entropy.  We also set the AdS radius to $\ell_{\rm AdS} = 1$.

According to the proposal of \cite{Ryu:2006bv,Hubeny:2007xt}, the holographic entanglement entropy of disjoint intervals in 2d CFTs is computed by the minimal length collection of geodesics that extend in the bulk and join the intervals endpoints on the boundary \cite{Headrick:2010zt}. 
To compute the geodesics associated to the offset intervals, we follow \cite{Hartman:2013qma} and exploit the local equivalence between the BTZ black hole and empty AdS$_3$. This allows to compute the length of geodesics in the black string geometry directly in terms of the length of AdS$_3$ geodesics anchored on the conformal boundary, which are just semicircles. 

As reviewed in \cite{Hartman:2013qma}, the different regions of the BTZ black string can be mapped to the Poincar\'e patch of AdS$_3$:
\be
ds^2 = \frac{1}{u^2}( -dy_0^2 +dy_1^2 + du^2)\,. 
\ee
In particular, the exterior metric of the black string
\be \label{eq:BTZext}
ds^2 = - \sinh^2\rho\, dt_1^2 + \cosh^2 \rho\, dx_1^2+ d\rho^2\,,
\ee
with conformal boundary at $\rho \to \infty$, maps to a portion of the Poincar\'e patch (we refer to the Appendix B in \cite{Hartman:2013qma} for the full explicit coordinates transformation).  Near the boundary, the two sets of coordinates are related by
\be \label{eq:bdrymap}
y_1\pm y_0 \approx e^{x_1\pm t_1}\,, \qquad  \frac{1}{u}  \approx \frac{1}{2} e^{\,\rho - x_1} \, .
\ee
These coordinates cover one of the exterior regions of the eternal BTZ string; the second exterior region with spatial coordinate $x_2$ is reached by continuing $t \to t + i \pi$. The boundaries of the two exterior regions thus cover the portion $y_0^2 - y_1^2 \le 0$ of the AdS boundary. 
In particular, a point $P$ on the right (left) boundary of the black sting in Poincar\'e coordinates reads 
\be
P=(y_0,y_1)= e^{x_{1,2}}(\sinh t_{1,2}, \pm \cosh t_{1,2} )\,.  
\ee 

Now, recall that the length of a spacelike geodesic in AdS$_3$ joining two points $P_1$, $P_2$ at radial coordinates $u_{P_1}$, $u_{P_2}$ near the AdS boundary $u=0$ is given by
\be \label{eq:LAdS}
\mathcal{L}_{12} = \log\left[ \frac{- \Delta y_0^2 + \Delta y _1^2}{ u_{P_1}u_{P_2} } \right]\,,
\ee
where $\Delta y_0 \equiv \(y_{0, P_1} - y_{0, P_2}\)$ and similarly for $\Delta y_1$. Using \eqref{eq:bdrymap}, it is then straightforward to obtain the length of spacelike geodesics anchored on the boundaries of the black string geometry. 

\begin{figure}
\centering
\includegraphics[width=0.6\textwidth]{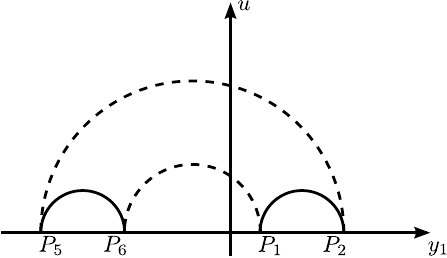} 
\caption{Illustration of the two competing configurations of geodesics for the offset interval in the $(y_1, u)$ plane. The disconnected configuration in solid lines is the one of minimal length for all times.}\label{fig:geodesics}
\end{figure}

The offset interval setup we are interested in corresponds to one interval of length $L$ on each boundary, shifted by a distance $D$, at some fixed time $t$. In Poincar\'e coordinates, this is the set of points:\footnote{
We use the same notation for the points as in sec.~\ref{sec:offset}.
}
\be \label{eq:points}
\begin{array}{lll}
P_1= (\sinh t,  \cosh t ) \, , &\quad & P_2= e^L (\sinh t, \cosh t ) \ , \\
P_5= e^{D+2L}(\sinh t, -  \cosh t ) \, , &\quad & P_6= e^{D+L}(\sinh t, -\cosh t )   \ .
\end{array}
\ee
In the computation of the holographic entanglement entropy, there are two competing sets of bulk geodesics ending on the endpoints of the offset interval. One consists of disconnected geodesics joining $P_1, P_2$ and $P_5, P_6$; while the second connected configuration of geodesics extends throughout the interior region of the BTZ string joining $P_1, P_6$ and $P_2, P_5$ (see fig.~\ref{fig:geodesics}). 

To regularize the length of these geodesics we introduce a radial cutoff $\rho_{\rm cutoff}$ in the coordinates \eqref{eq:BTZext} and define $ \eps \equiv e^{-\rho_{\rm cutoff}}/2$. In Poincar\'e coordinates, the cutoff is $u_{{\rm cutoff}1,2} =  \epsilon \, e^{x_{1,2}}$ and depends on the spatial coordinate $x_{1,2}$ of the point on the boundary. From \eqref{eq:LAdS} and \eqref{eq:points}, we then immediately obtain the entanglement entropy associated to the disconnected configuration
\be
S_{\rm disconnected}  = \frac{\mathcal{L}_{12} +\mathcal{L}_{56} }{ 4 G_{N} } = \frac{2c}{3} \log\left[  \frac{2\sinh (L/2)}{\eps}  \right] \,,
\ee
where we have used $c = 3/ (2 G_N)$. Reinserting the correct factors of $\beta$ and taking the scaling limit $L \gg \beta$, we have
\be
S_{\rm disconnected}  = 2S_0 +  \frac{2 \pi c L}{3 \beta} \,,
\ee
where, as in~\eqref{eq:S0}, we have isolated the divergent contribution
\be 
2S_0 = \frac{2c}{3} \log\frac{\beta}{2 \pi \epsilon}\, ,
\ee
For the second, connected, configuration we have instead
\be
S_{\rm connected}  = \frac{\mathcal{L}_{16} +\mathcal{L}_{25} }{ 4 G_{N} } = \frac{c}{3} \log\left[  2\frac{\cosh(2t) + \cosh(D+L) }{\eps^2} \right]\,.
\ee
Reinserting the correct factors of $\beta$ and taking $t ,D, L \gg \beta$, this reduces to
\bea
S_{\rm connected}  
&= & 2S_0 +  \frac{2 \pi c}{3 \beta}\left\{\begin{array}{ll}
D+L & \quad {\rm for} \quad \beta  < t < \frac{ D+L}{2} \\
2t & \quad {\rm for } \quad t > \frac{D+L}{2} 
\end{array}\right.\,.
\eea
Minimising over the different configurations then shows that, as expected, the disconnected configuration is always dominant, so the entanglement entropy is constant:
\be \label{HMoffset}
S=S_{\rm disconnected}  =  2S_0 + \frac{2 \pi c L }{3 \beta}  \,,
\ee
for all times. This corresponds to the `maximal scrambling' (dashed) line in figure \ref{fig:dip}, and to an identically vanishing mutual information. 

In particular, this result cannot be corrected by complex geodesics connecting real boundary endpoints, as in pure AdS these have the same length \eqref{eq:LAdS} of real ones.  

\subsection{Second R\'enyi entropy from semiclassical gravity}\label{ss:gravityren}

In section~\ref{sec:2ndRenyi} we studied the second R\'enyi entropy of the offset interval through the analysis of the torus partition function $Z(\tau, \bar \tau)$, and argued that generically the quasiparticle argument does not produce the correct 2nd R\'enyi. Here we compute the torus partition function from semiclassical gravity and explicitly verify that in holographic CFTs the quasiparticle dip in the second R\'enyi of the offset interval is suppressed by a factor 3. 

On the gravity side, the torus partition function is computed by a path integral over Euclidean three-geometries that are conformal at infinity to a torus of modular parameter $\tau = \frac{i\beta}{2\pi}$. The inverse temperature and angular momentum are respectively the real and imaginary parts of $\beta$. 

In the semiclassical limit, all known saddlepoints of pure 3d gravity that contribute to the partition function have classical action of the form \cite{Maldacena:1998bw,Maloney:2007ud}
\be \label{eq:Sgamma}
S(\tau, \bar \tau) = \frac{i \pi }{8 G_N}  (\gamma \cdot \tau - \gamma\cdot \bar \tau)\,,
\ee
with 
\be \label{eq:gamma}
\gamma\cdot \tau = \frac{a \tau +b}{c\tau +d}\,, \qquad \gamma = \left(\begin{array}{cc}a & b \\c & d\end{array}\right) \in SL(2,\mathbb Z)\,. 
\ee
Note that the same $SL(2,\mathbb Z)$ transformation acts on both $\tau$ and $\btau$.
The simplest saddlepoint is thermal AdS$_3$, with action
\be
S_{\rm thermal}(\tau, \bar \tau) =  \frac{i \pi }{8 G_N} (\tau -\bar \tau) \,,
\ee
which can be viewed as a solid torus with a contractible spatial cycle. Its $S$-transformation $\tau \to -\frac 1 \tau$ is the Euclidean BTZ black hole
\be
S_{\rm BTZ}(\tau, \bar \tau) = - \frac{i \pi }{8 G_N} \(\frac 1 \tau -\frac{1}{\bar \tau}\) \,.
\ee
The more general manifolds that can be constructed by a transformation $\gamma$ as in eq.~\eqref{eq:Sgamma}-\eqref{eq:gamma} form an $SL(2,\mathbb Z)$ family of black holes. 

The leading semiclassical approximation to the partition function is given by the solution of least action:
\be \label{eq:Zgrav}
\log Z(\tau, \bar \tau) \approx -S_{\rm min} (\tau, \bar \tau) = \frac{\pi }{8 G_N} {\rm max_{\gamma}}\, (- i \gamma \cdot \tau +i  \gamma \cdot \bar \tau)\,.
\ee
This leads to a rich phase diagram in the upper-half $\tau$ plane.
Thermal AdS space is the dominant classical solution in the fundamental domain $|\tau|>1, |{\rm Re}\, \tau| \le 1/2$, and in any of its translates by $\tau \to \tau+n$ for integer $n$, as the free energy is invariant under the $T$-transformation $\tau \to \tau+1$. Similarly, the Euclidean BTZ black hole dominates whenever there exists an integer $n$ such that $-\frac{1}{\tau+n}$ lies in the fundamental domain. The phase diagram resulting from \eqref{eq:Zgrav} is a subtesselation of the usual tesselation of the upper half $\tau$-plane by fundamental domains of $SL(2,\mathbb Z)$, with an infinite number of phases \cite{Maloney:2007ud}. For instance, for purely real $\beta$ one recovers the familiar Hawking-Page transition between thermal AdS at low temperatures and Euclidean BTZ at high temperatures. 

\begin{figure}
\centering
\includegraphics[width=5.75in]{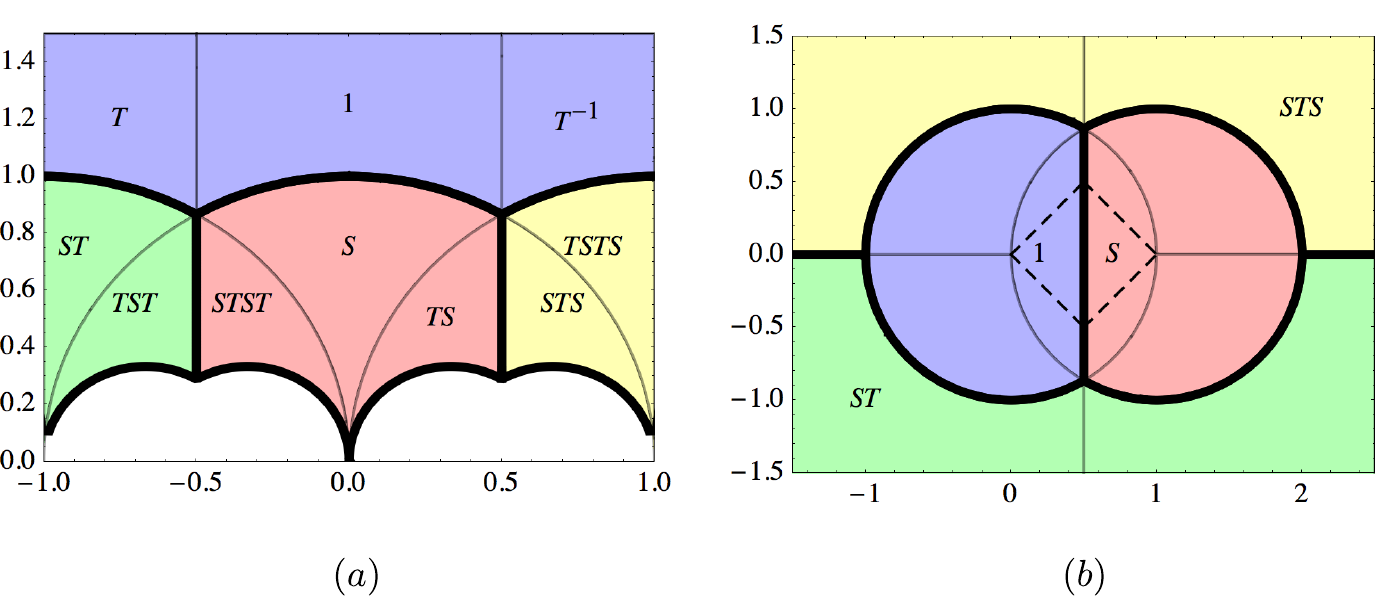}
\caption{$ (a)$ Phases of the partition function on the $\tau$ plane.  Regions are labeled by the element of $SL(2,\mathbb Z)$ which takes points in the region to the fundamental domain, marked `1'. Distinct phases are the four colored regions, separated by bold lines.  
$(b)$ Phases of $Z(\tau(x), \btau(x))$, on the $x$-plane.  Only the distinct phases $1, S, ST, STS$ are labeled.  The dashed diamond is responsible for the two phases in Lorentzian signature.\label{fig:phaseplots}}
\end{figure}

The modulus of the torus $\tau$ is related to the cross ratio $x$ in the four-point function that computes the second R\'enyi entropy of the offset interval through eq.~\eqref{xtau}
\be
\tau(x) = i \frac{K(1-x)}{K(x)} \ .
\ee
 The image of the $x$ plane under the map $x \to \tau(x)$ is shown in figure \ref{fig:phaseplots}$a$. Each region on the $\tau$ plane is labeled by the element of $SL(2,\mathbb Z)$ that takes this region to the fundamental domain.  This element also provides the value of the action in that region --- for example, in the region labeled $STS$, the action is proportional to the imaginary part of $S T S \cdot \tau = \frac{\tau}{1-\tau}$. Of all the regions shown, there are only four distinct values of the action, so these correspond to four different phases, shown in different colors.  The Euclidean action relevant to the twist correlator is the value of the action on the dominant saddle:
\be\label{logzz}
\log Z(x,\bx) \approx \frac{\pi}{8G_N}\max\left\{-i\tau + i \btau, \frac{i}{\tau} - \frac{i}{\btau}, -\frac{i}{1-\tau} + \frac{i}{1-\btau}, -\frac{i\tau}{1+\tau} + \frac{i\btau}{1+\btau}\right\}\ ,
\ee
where $\tau,\btau$ are functions of $x,\bx$.
The phase regions on the $x$ plane are shown in figure \ref{fig:phaseplots}$b$.

In Lorentzian signature the cross ratio varies inside the interval $[0,1]$.
This comes, upon analytic continuation, from the Euclidean diamond 
\be
{\rm Re}\, x > |{\rm Im} \, x|\,, \qquad 1-{\rm Re} \, x > |{\rm Im} \, x|\,
\ee
in the $x-$plane, shown as the dashed diamond in fig.~\ref{fig:phaseplots}$b$. As this Euclidean diamond intersects only the $\mathds{1}$ and $S$ phases, there is only one possible phase transition in Lorentzian signature:
\be
\log Z(x, \bar x) \approx  \frac{\pi c}{12} \, {\rm max}  \left\{ -i\tau + i \bar \tau \,, \quad \frac{i}{\tau} - \frac{i}{\bar \tau} \right\} \,. 
\ee
At early and late times ($t < D/2, t > D/2 +L $): $x, \bar x \sim 0$, the $\mathds{1}$ phase dominates and 
\be
\log Z(x, \bar x) \approx 2^{2 c/3} x^{-c/12} \bar x^{-c/12}
\ee
in complete agreement with \eqref{eq:ZlowT}. Substituting in the expression for the second R\'enyi \eqref{stgen}, we obtain
\be
S_2^{\rm early, late} = \frac{\pi c L}{2 \beta} +S_2^0\,.
\ee
At intermediate times ($D/2< t < D/2 +L$), we are in the mixed limit where the $\mathds{1}$ phase dominates again as $x \ll 1-\bar x \ll 1$. This gives 
\bea
\log Z(x, \bar x) &\sim& 2^{ c/3} x^{-c/12} \\
G_2 &\sim& 2^{ -c/3}x^{-c/8} (1-\bar x )^{-c/24}
\eea
and 
\be
S_2^{\rm dip} =  \frac{\pi c L}{2 \beta} - \frac{\pi c}{12 \beta} {\rm min} \(D+2L -2 t, 2t -D\) +S_2^0
\ee
in agreement with \eqref{eq:2ndRenyidip} for $c_{\rm currents} \ll c$. (We have dropped a term $\sim c \log 2$ which is subleading in the scaling limit $\beta \to 0$.)

Observe in particular that a light cone singularity that leads to a second R\'enyi in full agreement with the quasiparticle picture could only come from saddlepoints with action proportional to $\tau - \frac{1}{\bar \tau}$. These are the kind of holomorphically factorizing saddles that were sought, but not found, in \cite{Maloney:2007ud}. It was pointed out in \cite{Maloney:2007ud} that such saddlepoints would never dominate the classical action in Euclidean signature, but they can dominate in Lorentzian signature.


\subsection{Large-$c$ CFT}\label{sec:CFTlargec}

We will now discuss how these results are reproduced in large-$c$ CFT.

The universality class of CFTs with a semiclassical holographic dual is not characterized precisely, but it must satisfy at least two simple criteria (discussed for example in \cite{Heemskerk:2009pn,Hartman:2014oaa}).  First, these theories have $c \gg 1$.  The central charge is related to gravitational parameters by $c = 3 \ell_{\rm AdS}/(2G_N)$, where $\ell_{\rm AdS}$ is the anti-de Sitter radius and $G_N$ is the three-dimensional Newton's constant, so the semiclassical limit on the gravity side is the $c\to\infty$ limit in CFT.  Second, the spectrum of low-dimension operators must be sparse --- each low-dimension operator corresponds to a light field on the gravity side, and the number of such fields should be finite as $c \to \infty$. Exactly how to define the sparseness condition is not known (except for the partition function \cite{Hartman:2014oaa}), but we will assume that the spectrum is sufficiently sparse to suppress certain contributions to the conformal block expansion.

In section~\ref{sec:HM} we computed the holographic entanglement entropy of the offset interval in the thermal double model.  As discussed in \ref{sec:offset}, the entanglement and R\'enyi entropies are obtained in terms of the four-point twist correlator in the plane  $ G_n(x,\bx)$ in \eqref{gxx} via  
\be
S_n = \frac{1}{1-n} \log I^{\text{cylinder}}_n
\ee
with 
\be
I^{\text{cylinder}}_n = \(\frac{\beta}{\pi} \sinh\frac{\pi L}{\beta} \)^{-8 h_n} |x \bx |^{2h_{n}}  G_n(x,\bx)\,.
\ee
The general semiclassical conformal block analysis to evaluate $G_n(x,\bx)$  for CFTs with an holographic dual has been performed in \cite{Hartman:2013mia}.
There it was argued that, at leading order in $c$, the four-twist correlator is given by the dominant contribution between the $s$- and $t$-channel semiclassical identity contributions. That is, in these theories, the R\'enyi entropies are universal, with the twist correlator given by
\be\label{gfour}
G_n(x,\bx) = \max\left( F_0(cn,h_n; x)F_0(cn,h_n; \bx) , \ F_0(cn,h_n; 1-x)F_0(cn, h_n; 1-\bx)\right) \ ,
\ee
with $F_0$ the vacuum block for the algebra $\text{Vir}^n/\mathbb{Z}_n$.
For $n = 2$, a complete derivation of this formula with a suitable sparseness assumption is given in \cite{Hartman:2014oaa}, and the results match exactly with the gravity analysis of the 2nd R\'enyi in section \ref{ss:gravityren}.  For general $n$, additional caveats apply to \eqref{gfour}, as discussed in \cite{Hartman:2013mia}.  Assuming that it is correct, the right-hand side can be evaluated in the limit $n \to 1$ following \cite{Hartman:2013mia}, with the result that only the identity operator contributes:
\be
G_n(x,\bx) \approx [\min\{x \bx, (1-x)(1-\bx)\}]^{-\frac{c}{6}(n-1)} \qquad (n\to 1) \ .
\ee
There is no quasiparticle singularity.

Applying this result to the computation of the entanglement entropy of the offset intervals, one immediately recovers the analysis of the previous section and the holographic result \eqref{HMoffset}. For all times,  in the scaling limit $L \gg \beta$, $G_n(x,\bx)$ is given by the $s$-channel semiclassical identity block, resulting in the constant entanglement entropy 
\be
S= 2S_0 + \frac{2 \pi c L }{3 \beta}  \, .
\ee

A similar analysis applies to the 8-point function in the thermal double model for the quench.  Once again the $n\to 1$ limit of the twist correlator in a large-$c$ CFT is given by maximizing the identity contribution over OPE channels.  Each OPE channel corresponds to a choice of Ryu-Takayanagi geodesics connecting the 8 twist insertions, so this CFT result is identical to the gravity prediction.


\subsection{The D1--D5 CFT} \label{ss:stringy}

A specific realization of $\AdS_3$/$\CFT_2$ duality, and hence an in-principle 
non-perturbative formulation of quantum gravity, is afforded by the D1--D5 CFT. We will use this example to illustrate how memory effects in the entanglement depend on $N$, $\beta$, and the CFT coupling constant.

This system is realized in type IIB string theory compactified on a circle $S^1$ times $M^4$, 
with $M^4$ either a four-torus or a K3 surface, 
with $N_1$ D1-branes wrapped around the circle and $N_5$ D5-branes wrapped around the 
entire compact product space. The near horizon geometry is $\AdS_3 \times S^3 \times M^4$, 
and one can formulate a two-dimensional CFT at the conformal boundary of the AdS$_3$.
This CFT has central charge $c = 6N_1 N_5$ and $\mathcal{N} = (4,4)$ supersymmetry. (For reviews, see \cite{David:2002wn, Avery:2010qw}.)

The moduli space of the CFT is twenty-dimensional and includes a special point known as the 
``orbifold point." In the torus case, the theory at this point is a symmetric orbifold CFT with target 
space $(T^4)^{N_1 N_5} / S_{N_1 N_5}$. This is analogous to free super Yang-Mills theory in the 
$\AdS_5/\CFT_4$ duality, and the CFT at this point is $N\equiv N_1 N_5$ 
copies of a $c=6$ free CFT (the seed theory), composed of 4 real bosons and their fermionic superpartners, 
with an $S_N$ orbifold permuting the $N$ copies.
Though built out of free theories, the symmetric orbifold structure couples the various 
conformal families of the seed theories so that at large $N$ the quasiparticle picture 
of entanglement propagation is qualitatively altered, as we see in the following subsection.

\subsubsection*{Free theory}

We will focus on the second R\'enyi entropy of the offset intervals in the thermal double, since, as discussed in section \ref{sec:2ndRenyi}, this is fixed by the torus partition function.  The spectrum of the orbifold is known exactly \cite{Dijkgraaf:1996xw}, and in appendix \ref{app:zorb}, we extract from this the vacuum contribution to the torus partition function as $q \to 0, \bq \to 1$:
\be\label{zorb}
Z^{\text{orbifold}}(q,\bq) \approx \frac{1}{N!} q^{-c/24} (\bq')^{-c/24}
\ee
The limit leading to \eqref{zorb} is taken with $N$ held fixed. This theory has a very large number of conserved currents; it is current dominated and $c_{\rm currents} = c$. However, the orbifolding leads to the $1/N!$ prefactor, which means there are far fewer currents at large $N$ than in a simple product of free theories.

In the scaling limit $\beta \to 0$ with $N$ held fixed, we can ignore the coefficient as we did in section~\ref{sec:2ndRenyi}, and the orbifold theory exhibits the quasiparticle dip in the 2nd R\'enyi entropy. On the other hand, the coefficient of the quasiparticle singularity in \eqref{zorb} is suppressed by $N! \sim e^{N\log N}$.  Therefore to see the quasiparticle behavior, we must take
\be\label{logn}
\frac{t}{\beta}, \frac{L}{\beta}, \frac{D}{\beta} \gg  \log c \ .
\ee
At shorter times, the large coefficient suppresses the quasiparticle dip and \eqref{zorb} may not be the dominant term.

The free CFT is believed to be dual to a string theory with vanishing string tension (see for example \cite{Gaberdiel:2014cha,Gaberdiel:2015mra}). It would be interesting to interpret \eqref{logn} in the string theory. Curiously, the time scale $\beta \log c$ is the scrambling time in the gravity limit, but here plays a different role.

\subsubsection*{Deformed theory}

Semiclassical gravity is far in moduli space from the free orbifold CFT.  Roughly speaking, the coupling constant in the CFT is an exactly marginal deformation that corresponds to the string tension, and gravity is a good approximation when the string tension is very large. 

Little is known about how to follow this deformation in general, though many quantities protected by supersymmetry can be matched in the two limits.  To understand the behavior of the second R\'enyi entropy, we are interested in the spectrum of conserved currents as a function of the coupling constant.  In the free limit, it is given by \eqref{zorb}.  In the gravity limit, conserved currents correspond to massless gauge fields in supergravity, and the only such fields are the graviton, gravitino, and gauge fields required by $(4,4)$ supersymmetry.  Thus all other conserved currents contributing to \eqref{zorb}, with dimension $(0,\bh)$ at zero coupling, must pick up a non-zero left-moving conformal weight when we move to the gravity point in moduli space. 

It seems likely that all of these currents are lifted even at leading order in the deformation away from the free orbifold point, since they are not protected. This was recently confirmed explicitly for the low-lying currents using conformal perturbation theory \cite{Gaberdiel:2015uca}. The leading deformation, which has been studied perturbatively for example in \cite{Avery:2010er, Asplund:2011cq, Burrington:2012yq,Burrington:2012yn,Carson:2014ena}, is by an $S_N$-twist operator, so it exists for any $N \geq 2$. Therefore, we expect the CFT defined by a small deformation of the orbifold theory with $c \geq 12$ to be a tractable example of a CFT with $c_{\rm currents}<c$. 


\section{Conclusion} \label{s:discussion}

We have argued that there is a qualitative difference in the entanglement properties of theories with $c_{\rm currents} = c$ and $c_{\rm currents} < c$. Although CFTs with fewer conserved currents seem more generic in some sense, most of the well known CFTs in 1+1 dimensions are rational and therefore $c_{\rm currents} = c$.  What is the simplest theory with $c_{\rm currents} < c$? We do not have an answer to this question, but will suggest two candidates.\footnote{We thank Leonardo Rastelli for discussions about this question.} 

The first natural candidate is a theory with several scalars, and an arbitrary quartic potential. If the quartic potential is $O(N)$ invariant then the fixed points are well studied, but for a generic potential, the symmetry is reduced.  If such a theory has a fixed point with reduced symmetry then it seems likely to have $c_{\rm currents} < c$.  

Another class of well known examples are supersymmetric sigma models with a Calabi-Yau target space.  A generic Calabi-Yau has no isometries, so the sigma model has only the conserved currents required by superconformal symmetry.  A simple example of this type is the minimal version of the D1-D5 CFT discussed in section \ref{ss:stringy}. It is a deformation of the orbifold theory with target $(T^4)^2/\mathbb{Z}_2$.  This theory has central charge $c=12$. The ${\cal N}=(4,4)$ superconformal algebra consists of the stress tensor, four $R$-currents, and four supercurrents, so assuming that other currents are lifted by the deformation, $c_{\rm currents} \leq 9 < c$ and it follows that the entanglement entropy differs from the quasiparticle picture. 

Another question left open by our analysis is how to compute $S_{A\cup B}(t)$ in theories that have $c_{\rm currents} < c < \infty$.  We have argued that, in the dip regime, the entanglement entropy differs from the quasiparticle value, since the R\'enyi entropies disagree with the quasiparticle prediction.  
It would be very interesting to compute the vacuum contribution explicitly, which is the universal answer in the holographic limit. For theories with no extended algebra it depends only on the central charge. 

The suppression or absence of the dip in the two-intervals entanglement entropy after a quench might at first seem to clash with holographic computations for $D<L$ \cite{Asplund:2013zba,Balasubramanian:2011at,Allais:2011ys}. In fact, for small separation between the intervals, the holographic mutual information after a quench has a spike (or a bump in the case of a local quench) at early times, which seems in qualitative agreement with the quasiparticle picture  \cite{Asplund:2013zba,Balasubramanian:2011at,Allais:2011ys}.  This spike though has a different origin than the dip, and can be traced in the growth of $S_{A \cup B}$ to saturation. For $D<L$, $S_{A \cup B}$ starts growing as $2t$ for $t<D/2$, and continues $\sim t$ till saturation at time $L - D/2$. When combined with the growth $\sim t$ till $t=L/2$ of the single interval entanglement entropies $S_{A}, S_{B}$, this effect produces a spike between $D/2 < t < L-D/2$ in the mutual information.

In the main text, we have discussed extensively global quenches in the boundary state \cite{Calabrese:2005in,Calabrese:2007rg,Calabrese:2009qy} and thermal double model \cite{Hartman:2013qma}, but similar analysis and conclusions hold for the local `joining' quenches discussed in \cite{localquench,Asplund:2013zba}. These describe the process of two semi-infinite lines joined at their endpoints at an instant of time and subsequently evolving as a connected, infinite system. The computation of the two-intervals entanglement and R\'enyi entropies in these systems can be reduced to that of a four-point function of twist operators in the UHP and thus proceeds in an analogous way to that presented in section~\ref{s:origin}. 

Closely related systems are the heavy local operator quenches considered in \cite{Nozaki:2013wia, Asplund:2013zba, Asplund:2014coa,Caputa:2015waa}. These are systems quenched by the insertion of a heavy local primary with weights $h, \bar h \sim c$ that produces a localized excitation. 
The operator quench with $h =\bar h  =c/32$ has the same energy-momentum tensor for $t>0$ as the Calabrese-Cardy `joining' quench \cite{localquench} and, for intervals sufficiently distant from the quench, exhibits the same evolution for the entanglement entropy.
Therefore, the same observations and conclusions  apply, and, in particular, our results resolve the apparent discrepancy between the quasiparticle picture and the holographic calculations observed in \cite{Asplund:2013zba}.
Moreover, in the limit of large-$c$, the absence of the dip in $S_{A\cup B}$ after a heavy local operator quench can be immediately inferred from the results of \cite{Fitzpatrick:2015zha} for the six-point function of two heavy and four light operators. 

The implications of our methods might also be relevant to the light operator quenches studied in \cite{Nozaki:2014hna,He:2014mwa,Caputa:2014vaa,Nozaki:2014uaa,Caputa:2014eta}.
Similarly, we expect our observations on light cone singularities to be important in the computation of the entanglement negativity of two disjoint intervals $A, B$ after a global or local quench \cite{Zimboras,Coser:2014gsa,Hoogeveen:2014bqa,Wen:2015qwa}, which in \cite{Coser:2014gsa,Wen:2015qwa} was found to be exactly 3/4 of the quasiparticle mutual information $I (A,B)$.

\bigskip

\bigskip

\textbf{Acknowledgments}
We are grateful to Steven Avery, Ben Burrington, John Cardy, Frederik Denef, Liam Fitzpatrick, Veronica Hubeny, Jared Kaplan, Edoardo Lauria, Carlo Maccaferri, Alex Maloney, Don Marolf, Amanda Peet, Mukund Rangamani, Leonardo Rastelli, David Simmons-Duffin, Erik Tonni, Aron Wall, and Ida Zadeh for useful discussions.  
CA is supported in part by a grant from the John Templeton Foundation and in part by the U.S.~Department of Energy under DOE grant DE-FG02-92-ER40699. The opinions expressed in this publication are those of the authors and do not necessarily reflect the views of the John Templeton Foundation. 
The work of AB and FG is supported in part by the Belgian Federal Science Policy Office through the Interuniversity Attraction Pole P7/37, by FWO-Vlaanderen through project G.0651.11, by the COST Action MP1210 The String Theory Universe and by the European Science Foundation Holograv Network. AB and FG are FWO-Vlaanderen postdocs. AB and FG also thank the Galileo Galilei Institute for Theoretical Physics for the hospitality and the INFN for partial support during a portion of this work. 
TH is supported by a DOE early career grant, and also acknowledges support from the KITP under NSF PHY11-25915 during the program ``Entanglement in Strongly-Correlated Quantum Matter.''

\bigskip

\bigskip

\appendix

\section{Vacuum character of the symmetric orbifold}\label{app:zorb}
In this appendix we calculate the vacuum character of the symmetric orbifold CFT with target space $(T^4)^N/S_N$, following \cite{Dijkgraaf:1996xw,Gaberdiel:2014cha,Gaberdiel:2015mra}, and find the asymptotics claimed in section \ref{ss:stringy}. 

The theory on $T^4$ has four free bosons and four free fermions, and therefore central charge 6.  The vacuum character of this seed theory in the R sector is
\bea
\chi_\text{R}(q,y) &=& \sum_{n=0}^\infty \sum_{\ell \in \mathbb{Z}} c(n,\ell)q^n y^\ell\\
&=& (y-2+y^{-1})\prod_{n=1}^\infty \frac{(1-yq^n)^2(1-y^{-1}q^n)^2}{(1-q^n)^4}\, , 
\eea
where $y$ corresponds to the chemical potential.
The coefficients $c(n,\ell)$ defined by this expansion can be used to construct the generating function of the NS-sector vacuum characters in the orbifold theories with $c=6N$ \cite{Gaberdiel:2014cha,Gaberdiel:2015mra}:
\be\label{genfun}
\sum_{N=0}^\infty p^N \chi^{\text{vac}}_N(q) = \prod_{n=0}^\infty \prod_{\ell \in \mathbb{Z}}(1+(-1)^\ell p q^{n+\ell/2+1/4})^{-c(n,\ell)} \ , 
\ee
where we have set $y=1$.
Our goal is to find $\chi^{\text{vac}}_N(q)$ as $q\to 1$.

Define the NS vacuum character of the seed theory  (Tr$_{\text{NS}}q^{L_0-c/24}$),
\bea
\chi(q) &=& \sum_{n=0}^\infty \sum_{\ell \in \mathbb{Z}}c(n,\ell)q^{n+\ell/2+1/4}(-1)^{\ell+1}\\
&=& \frac{\theta_3(\tau)^2}{\eta(\tau)^6}  \\
&=& q^{-1/4}(1 + 4 q^{1/2} + 10 q + 24 q^{3/2} + 55 q^2 + \cdots )
\eea
and the alternating character (Tr$_{\text{NS}}(-1)^F q^{L_0-c/24}$):
\bea
\tilde{\chi}(q) &=& \sum_{n=0}^\infty \sum_{\ell \in  \mathbb{Z}} c(n ,\ell)q^{n+\ell/2+1/4}\\
 &=& \frac{\theta_4(\tau)^2}{\eta(\tau)^6}   \\
&=& q^{-1/4}(1 - 4 q^{1/2} + 10 q-24q^{3/2} +55 q^2 + \cdots)
\eea
Using these we can re-express the generating function \eqref{genfun} as
\be\label{genfun2}
\sum_{N=0}^\infty p^N \chi^{\text{vac}}_N(q) = \exp\left[ \sum_{k=1,3,\dots} \frac{p^k}{k} \chi(q^k) + \sum_{k=2,4,\dots}\frac{p^k}{k}\tilde{\chi}(q^k)\right]
\ee
As $q \to 1^-$, 
\be\label{chisim}
\chi(q) \sim -\tau^2 (q')^{-1/4} ,\qquad \tilde{\chi}(q) \sim -4\tau^2 \ ,
\ee
and
\be
\chi(q^k) \sim -(k\tau)^2 e^{i \pi /(2k\tau)} \ ,
\ee
where $q' = e^{-2\pi i/\tau}$. The first equation in \eqref{chisim} indicates that the seed theory has $c_{\rm currents} = c = 6$.

We are computing the limit $q \to 1^-$ with $N$ held fixed, so we can plug these asymptotics into \eqref{genfun2}. The leading singularity comes from $k=1$:
\be\label{chifinal}
\chi_N^{\text{vac}}(q) \sim \frac{1}{N!} (q')^{-N/4} \ .
\ee
This is the result \eqref{zorb} used in the discussion of the second R\'enyi. The character \eqref{chifinal} is Tr$_{\text{NS}}q^{L_0-c/24}$ in the orbifold. The full partition function $Z(\tau,\btau)$ includes a sum over spin structures, but the other terms are subleading in this limit.
%


\end{spacing}


\begin{thebibliography}{99}%
\small

\bibitem{bloch}
  M.~Cheneau, P.~Barmettler, D.~Poletti, M.~Endres, P.~Schauss, T.~Fukuhara, C.~Gross, I.~Bloch, C.~Kollath, and S.~Kuhr,
  ``Light-cone-like spreading of correlations in a quantum many-body system,''
 Nature {\bf 481}, 484177 (2012) 
 [\arXiv{arXiv:1111.0776} [cond-mat.quant-gas]] 
 
T.~Langen, R.~Geiger, and J.~Schmiedmayer, 
 ``Ultracold atoms out of equilibrium," 
 Annual Review of Condensed Matter Physics, {\bf 6}, (2015) 201-217
[\arXiv{arXiv:1408.6377} [cond-mat.quant-gas]].

\bibitem{cm}
  J.~Eisert and T.~J.~Osborne,
  ``General Entanglement Scaling Laws from Time Evolution,''
  Phys.\ Rev.\ Lett.\  {\bf 97} (2006) 150404
  [\arXiv{arXiv:quant-ph/0603114}]. 

J.~H.~Bardarson, F.~Pollmann, and J.~E.~Moore, ``Unbounded growth of entanglement in models of many-body localization,''
Phys.~Rev.~Lett.~ 109, 017202 (2012), 
[\arXiv{arXiv:1202.5532} [cond-mat.str-el]].

\bibitem{Sekino:2008he} 
  Y.~Sekino and L.~Susskind,
  ``Fast Scramblers,''
  JHEP {\bf 0810}, 065 (2008)
  [\arXiv{arXiv:0808.2096} [hep-th]].

M.~Rigol and M.~Srednicki, 
``Alternatives to Eigenstate Thermalization,''
 Phys.\ Rev.\ Lett.\  {\bf 108} 110601 (2011)
[\arXiv{arXiv:1108.0928} [cond-mat.stat-mech]].

  J.~Maldacena, S.~H.~Shenker and D.~Stanford,
  ``A bound on chaos,''
 [\arXiv{arXiv:1503.01409} [hep-th]].
  
\bibitem{holoq}

  J.~Abajo-Arrastia, J.~Aparicio and E.~Lopez,
  ``Holographic Evolution of Entanglement Entropy,''
  JHEP {\bf 1011}, 149 (2010)
  [\arXiv{arXiv:1006.4090} [hep-th]].
  
  T.~Albash and C.~V.~Johnson,
  ``Evolution of Holographic Entanglement Entropy after Thermal and Electromagnetic Quenches,''
  New J.\ Phys.\  {\bf 13}, 045017 (2011)
  [\arXiv{arXiv:1008.3027} [hep-th]].

V.~Balasubramanian, A.~Bernamonti, J.~de Boer, N.~Copland, B.~Craps, E.~Keski-Vakkuri, B.~Muller and A.~Schafer {\it et al.},
  ``Thermalization of Strongly Coupled Field Theories,''
  Phys.\ Rev.\ Lett.\  {\bf 106}, 191601 (2011)
  [\arXiv{arXiv:1012.4753} [hep-th]]. 
  
  H.~Liu and S.~J.~Suh,
  ``Entanglement Tsunami: Universal Scaling in Holographic Thermalization,''
  Phys.\ Rev.\ Lett.\  {\bf 112}, 011601 (2014)
  [\arXiv{arXiv:1305.7244} [hep-th]].  
  
  S.~H.~Shenker and D.~Stanford,
  ``Black holes and the butterfly effect,''
  JHEP {\bf 1403} (2014) 067
  [\arXiv{arXiv:1306.0622 } [hep-th]]. 
     
    V.~E.~Hubeny and H.~Maxfield,
  ``Holographic probes of collapsing black holes,''
  JHEP {\bf 1403}, 097 (2014)
  [\arXiv{arXiv:1312.6887} [hep-th]].
  
\bibitem{Hartman:2013qma}
  T.~Hartman and J.~Maldacena,
  ``Time Evolution of Entanglement Entropy from Black Hole Interiors,''
  JHEP {\bf 1305} (2013) 014
  [\arXiv{arXiv:1303.1080 } [hep-th]].
    
\bibitem{Calabrese:2005in}
  P.~Calabrese and J.~L.~Cardy,
  ``Evolution of entanglement entropy in one-dimensional systems,''
  J.\ Stat.\ Mech.\  {\bf 0504} (2005) P04010
  [\arXiv{cond-mat/0503393}].    
    
\bibitem{Calabrese:2007rg}
  P.~Calabrese and J.~Cardy,
  ``Quantum Quenches in Extended Systems,''
  J.\ Stat.\ Mech.\  {\bf 0706} (2007) P06008
  [\arXiv{arXiv:0704.1880} [cond-mat.stat-mech]].  
    
\bibitem{Calabrese:2009qy}
  P.~Calabrese and J.~Cardy,
  ``Entanglement entropy and conformal field theory,''
  J.\ Phys.\ A {\bf 42} (2009) 504005
  [\arXiv{arXiv:0905.4013} [cond-mat.stat-mech]]. 
  
\bibitem{MIbound}  
M.~M.~Wolf, F.~Verstraete, M.~B.~Hastings, and J.~I.~Cirac, 
``Area laws in quantum systems: Mutual information and correlations,'' 
Phys.\ Rev.\ Lett.\ {\bf 100} (Feb, 2008) 070502 [\arXiv{arXiv:0704.3906} [quant-ph]].  
  
\bibitem{Asplund:2013zba}
  C.~T.~Asplund and A.~Bernamonti,
  ``Mutual information after a local quench in conformal field theory,''
  Phys.\ Rev.\ D {\bf 89} (2014) 6,  066015
  [\arXiv{arXiv:1311.4173} [hep-th]].
  
\bibitem{Balasubramanian:2011at}
  V.~Balasubramanian, A.~Bernamonti, N.~Copland, B.~Craps and F.~Galli,
  ``Thermalization of mutual and tripartite information in strongly coupled two dimensional conformal field theories,''
  Phys.\ Rev.\ D {\bf 84} (2011) 105017
  [\arXiv{arXiv:1110.0488} [hep-th]].  
  
\bibitem{Allais:2011ys}
  A.~Allais and E.~Tonni,
  ``Holographic evolution of the mutual information,''
  JHEP {\bf 1201} (2012) 102
  [\arXiv{arXiv:1110.1607} [hep-th]].  
    
\bibitem{Leichenauer:2015xra}
  S.~Leichenauer and M.~Moosa,
 ``Entanglement Tsunami in (1+1)-Dimensions,''
   \arXiv{arXiv:1505.04225} [hep-th].
    
\bibitem{Coser:2014gsa}
  A.~Coser, E.~Tonni and P.~Calabrese,
  ``Entanglement negativity after a global quantum quench,''
  J.\ Stat.\ Mech.\  {\bf 1412} (2014) 12,  P12017
  [\arXiv{arXiv:1410.0900} [cond-mat.stat-mech]].
  
\bibitem{Calabrese:2006rx}
  P.~Calabrese and J.~L.~Cardy,
  ``Time-dependence of correlation functions following a quantum quench,''
  Phys.\ Rev.\ Lett.\  {\bf 96} (2006) 136801
  [\arXiv{cond-mat/0601225}].
 
\bibitem{Holzhey:1994we}
  C.~Holzhey, F.~Larsen and F.~Wilczek,
  ``Geometric and renormalized entropy in conformal field theory,''
  Nucl.\ Phys.\ B {\bf 424} (1994) 443
  [\arXiv{hep-th/9403108}].
 
\bibitem{Calabrese:2004eu}
  P.~Calabrese and J.~L.~Cardy,
  ``Entanglement entropy and quantum field theory,''
  J.\ Stat.\ Mech.\  {\bf 0406} (2004) P06002
  [\arXiv{hep-th/0405152}].
 
 \bibitem{BCFT}
 J.~Cardy, 
 ``Conformal invariance and surface critical behavior,'' 
 Nucl.\ Phys.\ B {\bf 240} (1984) 514.

\bibitem{Headrick:2010zt}
  M.~Headrick,
  ``Entanglement Renyi entropies in holographic theories,''
  Phys.\ Rev.\ D {\bf 82} (2010) 126010
  [\arXiv{arXiv:1006.0047} [hep-th]].
  
\bibitem{Lunin:2000yv} 
  O.~Lunin and S.~D.~Mathur,
  ``Correlation functions for M**N / S(N) orbifolds,''
  Commun.\ Math.\ Phys.\  {\bf 219}, 399 (2001)
  [\arXiv{hep-th/0006196}].
  
\bibitem{Cardy:1986ie} 
  J.~L.~Cardy,
  ``Operator Content of Two-Dimensional Conformally Invariant Theories,''
  Nucl.\ Phys.\ B {\bf 270}, 186 (1986).
  
\bibitem{Zamolodchikov:2001ah} 
  A.~B.~Zamolodchikov and A.~B.~Zamolodchikov,
  ``Liouville field theory on a pseudosphere,''
  \arXiv{hep-th/0101152}.
  
\bibitem{Kitaev:2005dm} 
  A.~Kitaev and J.~Preskill,
  ``Topological entanglement entropy,''
  Phys.\ Rev.\ Lett.\  {\bf 96}, 110404 (2006)
  [\arXiv{hep-th/0510092}].
  
\bibitem{Levin:2006zz} 
  M.~Levin and X.~G.~Wen,
  ``Detecting Topological Order in a Ground State Wave Function,''
  Phys.\ Rev.\ Lett.\  {\bf 96}, 110405 (2006) [\arXiv{arXiv:cond-mat/0510613} [cond-mat.str-el]].
  
\bibitem{He:2014mwa}
  S.~He, T.~Numasawa, T.~Takayanagi and K.~Watanabe,
  ``Quantum dimension as entanglement entropy in two dimensional conformal field theories,''
  Phys.\ Rev.\ D {\bf 90} (2014) 4,  041701
  [\arXiv{arXiv:1403.0702} [hep-th]].
 
\bibitem{Pappadopulo:2012jk} 
  D.~Pappadopulo, S.~Rychkov, J.~Espin and R.~Rattazzi,
  ``OPE Convergence in Conformal Field Theory,''
  Phys.\ Rev.\ D {\bf 86}, 105043 (2012)
  [\arXiv{arXiv:1208.6449} [hep-th]].
  
\bibitem{zamof}
A.~B.~Zamolodchikov, 
``Two-dimensional conformal symmetry and critical four-spin correlation functions in the Ashkin-Teller model," Sov. Phys. - JETP {\bf 63} 1061 (1986).
    
\bibitem{Runkel:2001ng} 
  I.~Runkel and G.~M.~T.~Watts,
  ``A Nonrational CFT with c = 1 as a limit of minimal models,''
  JHEP {\bf 0109}, 006 (2001)
  [\arXiv{hep-th/0107118}].
    
\bibitem{Ginsparg:1988ui} 
  P.~H.~Ginsparg,
  ``Applied Conformal Field Theory,''
  \arXiv{hep-th/9108028}.

\bibitem{DiFrancesco1997} 
  P.~Di Francesco, P.~Mathieu and D.~Senechal,
  \textit{Conformal Field Theory},
  Springer (1997).

\bibitem{Dijkgraaf:1987vp} 
  R.~Dijkgraaf, E.~P.~Verlinde and H.~L.~Verlinde,
  ``C = 1 Conformal Field Theories on Riemann Surfaces,''
  Commun.\ Math.\ Phys.\  {\bf 115}, 649 (1988).
  
   \bibitem{Halpern:1995js}
  M.~B.~Halpern, E.~Kiritsis, N.~A.~Obers and K.~Clubok,
  ``Irrational conformal field theory,''
  Phys.\ Rept.\  {\bf 265} (1996) 1
  [\arXiv{hep-th/9501144}].
  
\bibitem{Calabrese:2009ez} 
  P.~Calabrese, J.~Cardy and E.~Tonni,
  ``Entanglement entropy of two disjoint intervals in conformal field theory,''
  J.\ Stat.\ Mech.\  {\bf 0911}, P11001 (2009)
  [\arXiv{arXiv:0905.2069} [hep-th]].
  
\bibitem{Calabrese:2010he} 
  P.~Calabrese, J.~Cardy and E.~Tonni,
  ``Entanglement entropy of two disjoint intervals in conformal field theory II,''
  J.\ Stat.\ Mech.\  {\bf 1101}, P01021 (2011)
  [\arXiv{arXiv:1011.5482} [hep-th]].
  
\bibitem{neg}
P.~Calabrese, J.~Cardy, E.~Tonni,
``Entanglement negativity in extended systems: A field theoretical approach,"
J.~Stat.~Mech.~(2013) P02008
[\arXiv{arXiv:1210.5359} [cond-mat.stat-mech]].

\bibitem{Hayden:2011ag}
  P.~Hayden, M.~Headrick and A.~Maloney,
  ``Holographic Mutual Information is Monogamous,''
  Phys.\ Rev.\ D {\bf 87} (2013) 4,  046003
  [\arXiv{arXiv:1107.2940} [hep-th]].
  
\bibitem{Wall:2012uf}
  A.~C.~Wall,
  ``Maximin Surfaces, and the Strong Subadditivity of the Covariant Holographic Entanglement Entropy,''
  Class.\ Quant.\ Grav.\  {\bf 31} (2014) 22,  225007
  [\arXiv{arXiv:1211.3494} [hep-th]].
    
\bibitem{Caputa:2014vaa} 
  P.~Caputa, M.~Nozaki and T.~Takayanagi,
  ``Entanglement of local operators in large-N conformal field theories,''
  PTEP {\bf 2014}, no. 9, 093B06 (2014)
  [\arXiv{arXiv:1405.5946} [hep-th]].

\bibitem{Ryu:2006bv}
  S.~Ryu and T.~Takayanagi,
  ``Holographic derivation of entanglement entropy from AdS/CFT,''
  Phys.\ Rev.\ Lett.\  {\bf 96} (2006) 181602
  [\arXiv{hep-th/0603001}].
  
\bibitem{Hubeny:2007xt}
  V.~E.~Hubeny, M.~Rangamani and T.~Takayanagi,
  ``A Covariant holographic entanglement entropy proposal,''
  JHEP {\bf 0707} (2007) 062
  [\arXiv{arXiv:0705.0016} [hep-th]].
 
\bibitem{Maldacena:1998bw}
  J.~M.~Maldacena and A.~Strominger,
  ``AdS(3) black holes and a stringy exclusion principle,''
  JHEP {\bf 9812} (1998) 005
  [\arXiv{hep-th/9804085}].
 
\bibitem{Maloney:2007ud}
  A.~Maloney and E.~Witten,
  ``Quantum Gravity Partition Functions in Three Dimensions,''
  JHEP {\bf 1002} (2010) 029
  [\arXiv{arXiv:0712.0155} [hep-th]].
  
\bibitem{Heemskerk:2009pn} 
  I.~Heemskerk, J.~Penedones, J.~Polchinski and J.~Sully,
  ``Holography from Conformal Field Theory,''
  JHEP {\bf 0910}, 079 (2009)
  [\arXiv{arXiv:0907.0151} [hep-th]].
  
\bibitem{Hartman:2014oaa} 
  T.~Hartman, C.~A.~Keller and B.~Stoica,
  ``Universal Spectrum of 2d Conformal Field Theory in the Large c Limit,''
  JHEP {\bf 1409}, 118 (2014)
  [\arXiv{arXiv:1405.5137} [hep-th]].
   
\bibitem{Hartman:2013mia}
  T.~Hartman,
``Entanglement Entropy at Large Central Charge,''
  \arXiv{arXiv:1303.6955} [hep-th].

\bibitem{David:2002wn} 
  J.~R.~David, G.~Mandal and S.~R.~Wadia,
  ``Microscopic formulation of black holes in string theory,''
  Phys.\ Rept.\  {\bf 369}, 549 (2002)
  [\arXiv{hep-th/0203048}].
  
\bibitem{Avery:2010qw} 
  S.~G.~Avery,
  ``Using the D1D5 CFT to Understand Black Holes,''
  Ph.~D.~Dissertation, The Ohio State University (2010)
  [\arXiv{arXiv:1012.0072} [hep-th]].
  
\bibitem{Dijkgraaf:1996xw} 
  R.~Dijkgraaf, G.~W.~Moore, E.~P.~Verlinde and H.~L.~Verlinde,
  ``Elliptic genera of symmetric products and second quantized strings,''
  Commun.\ Math.\ Phys.\  {\bf 185}, 197 (1997)
  [\arXiv{hep-th/9608096}].
  
\bibitem{Gaberdiel:2014cha} 
  M.~R.~Gaberdiel and R.~Gopakumar,
  ``Higher Spins and Strings,''
  JHEP {\bf 1411}, 044 (2014)
  [\arXiv{arXiv:1406.6103} [hep-th]].
  
\bibitem{Gaberdiel:2015mra} 
  M.~R.~Gaberdiel and R.~Gopakumar,
  ``Stringy Symmetries and the Higher Spin Square,''
  J.\ Phys.\ A {\bf 48}, no. 18, 185402 (2015)
  [\arXiv{arXiv:1501.07236} [hep-th]].
  
\bibitem{Gaberdiel:2015uca} 
  M.~R.~Gaberdiel, C.~Peng and I.~G.~Zadeh,
  ``Higgsing the stringy higher spin symmetry,''
  \arXiv{arXiv:1506.02045} [hep-th].
  
\bibitem{Avery:2010er} 
  S.~G.~Avery, B.~D.~Chowdhury and S.~D.~Mathur,
  ``Deforming the D1D5 CFT away from the orbifold point,''
  JHEP {\bf 1006}, 031 (2010)
  [\arXiv{arXiv:1002.3132} [hep-th]].
  

  
\bibitem{Asplund:2011cq} 
  C.~T.~Asplund and S.~G.~Avery,
  ``Evolution of Entanglement Entropy in the D1-D5 Brane System,''
  Phys.\ Rev.\ D {\bf 84}, 124053 (2011)
  [\arXiv{arXiv:1108.2510} [hep-th]].
  
\bibitem{Burrington:2012yq} 
  B.~A.~Burrington, A.~W.~Peet and I.~G.~Zadeh,
  ``Operator mixing for string states in the D1-D5 CFT near the orbifold point,''
  Phys.\ Rev.\ D {\bf 87}, no. 10, 106001 (2013)
  [\arXiv{arXiv:1211.6699} [hep-th]].
  
\bibitem{Burrington:2012yn} 
  B.~A.~Burrington, A.~W.~Peet and I.~G.~Zadeh,
  ``Twist-nontwist correlators in $M^N/S_N$ orbifold CFTs,''
  Phys.\ Rev.\ D {\bf 87}, no. 10, 106008 (2013)
  [\arXiv{arXiv:1211.6689} [hep-th]].
  
\bibitem{Carson:2014ena} 
  Z.~Carson, S.~Hampton, S.~D.~Mathur and D.~Turton,
  ``Effect of the deformation operator in the D1D5 CFT,''
  JHEP {\bf 1501}, 071 (2015)
  [\arXiv{arXiv:1410.4543} [hep-th]].

\bibitem{localquench}
P.~Calabrese and J.~Cardy, 
``Entanglement and correlation functions following a local quench: a conformal field theory approach,'' 
J. Stat. Mech. (2007) P10004, [\arXiv{arXiv:0708.3750} [cond-mat.stat-mech]].  

\bibitem{Nozaki:2013wia}
  M.~Nozaki, T.~Numasawa and T.~Takayanagi,
  ``Holographic Local Quenches and Entanglement Density,''
  JHEP {\bf 1305} (2013) 080
  [\arXiv{arXiv:1302.5703} [hep-th]].

\bibitem{Asplund:2014coa}
  C.~T.~Asplund, A.~Bernamonti, F.~Galli and T.~Hartman,
  ``Holographic Entanglement Entropy from 2d CFT: Heavy States and Local Quenches,''
  JHEP {\bf 1502} (2015) 171
  [\arXiv{arXiv:1410.1392} [hep-th]].  

\bibitem{Caputa:2015waa}
  P.~Caputa, J.~Simon, A.~Stikonas, T.~Takayanagi and K.~Watanabe,
  ``Scrambling time from local perturbations of the eternal BTZ black hole,''
  \arXiv{arXiv:1503.08161} [hep-th].

\bibitem{Fitzpatrick:2015zha}
  A.~L.~Fitzpatrick, J.~Kaplan and M.~T.~Walters,
  ``Virasoro Conformal Blocks and Thermality from Classical Background Fields,''
  \arXiv{arXiv:1501.05315} [hep-th].
    
\bibitem{Nozaki:2014hna}
  M.~Nozaki, T.~Numasawa and T.~Takayanagi,
  ``Quantum Entanglement of Local Operators in Conformal Field Theories,''
  Phys.\ Rev.\ Lett.\  {\bf 112} (2014) 111602
  [\arXiv{arXiv:1401.0539} [hep-th]].
  
\bibitem{Nozaki:2014uaa}
  M.~Nozaki,
  ``Notes on Quantum Entanglement of Local Operators,''
  JHEP {\bf 1410} (2014) 147
  [\arXiv{arXiv:1405.5875} [hep-th]].
  
\bibitem{Caputa:2014eta}
  P.~Caputa, J.~Simon, A.~Stikonas and T.~Takayanagi,
 ``Quantum Entanglement of Localized Excited States at Finite Temperature,''
  JHEP {\bf 1501} (2015) 102
  [\arXiv{arXiv:1410.2287} [hep-th]].

\bibitem{Zimboras}
V.~Eisler and Z.~Zimbor\'as, 
``Entanglement negativity in the
harmonic chain out of equilibrium,'' 
New Journal of Physics, {\bf 16} (2014) 123020
[\arXiv{arXiv:1406.5474} [cond-mat.stat-mech]].

\bibitem{Hoogeveen:2014bqa}
  M.~Hoogeveen and B.~Doyon,
  ``Entanglement negativity and entropy in non-equilibrium conformal field theory,''
  Nucl.\ Phys.\ B {\bf 898} (2015) 78
  [\arXiv{arXiv:1412.7568} [cond-mat.stat-mech]].

\bibitem{Wen:2015qwa}
  X.~Wen, P.~Y.~Chang and S.~Ryu,
  ``Entanglement negativity after a local quantum quench in conformal field theories,''
  \arXiv{arXiv:1501.00568} [cond-mat.stat-mech].
 
 
\end{thebibliography}
\end{document}